\documentclass[fleqn,usenatbib]{mnras}

\usepackage{newtxtext,newtxmath}
\usepackage[T1]{fontenc}

\DeclareRobustCommand{\VAN}[3]{#2}
\let\VANthebibliography\thebibliography
\def\thebibliography{\DeclareRobustCommand{\VAN}[3]{##3}\VANthebibliography}

\usepackage{graphicx}	% Including figure files
\usepackage{amsmath}	% Advanced maths commands
\usepackage{caption}
\usepackage{subcaption}
\newcommand\nustar{{\it NuSTAR}}
\newcommand\xmm{{\it XMM-Newton}}

\newcommand\suz{{\it Suzaku}}
\newcommand\Swift{{\it Swift/XRT}}
\newcommand\swift{{\it Swift}}
\newcommand\s{{\rm~s}}

\newcommand\xiunit{\ifmmode {\rm~erg\s}$^{-1}$ \else ~erg~cm~s$^{-1}$\fi}
\newcommand\kms{\ifmmode {\rm~km\ s}$^{-1}$ \else ~km s$^{-1}$\fi}
\newcommand\Hunit{\ifmmode {\rm~km\ s}$^{-1}$\ {\rm Mpc}$^{-1}$
	\else ~km s$^{-1}$ Mpc$^{-1}$\fi}
\newcommand\cts{\ifmmode {\rm~count\ s}$^{-1}$ \else ~count s$^{-1}$\fi}
\newcommand\ergsec{\ifmmode {\rm~erg\ s}$^{-1}$ \else
	~erg s$^{-1}$\fi}
\newcommand\funit{\ifmmode {\rm~erg\ s}$^{-1}$\;{\rm cm}$^{-2}$ \else
	~erg s$^{-1}$ cm$^{-2}$\fi}
\newcommand\phflux{\ifmmode {\rm~photon\ s}$^{-1}$\;{\rm cm}$^{-2}$
	\else   ~photon s$^{-1}$ cm$^{-2}$\fi}
 \newcommand\normflux{\ifmmode {\rm~photons/keV};{\rm cm}$^{2}$
	\else   ~photons/keV/cm$^{2}$/s\fi}
\newcommand\efluxA{\ifmmode {\rm~erg\ s}$^{-1}$\;{\rm cm}$^{-2}$\;{\rm
		\AA}$^{-1}$ \else ~erg s$^{-1}$ cm$^{-2}$ \AA$^{-1}$\fi}
\newcommand\efluxHz{\ifmmode {\rm~erg\ s}$^{-1}$\;{\rm cm}$^{-2}$\;{\rm
		Hz}$^{-1}$ \else ~erg s$^{-1}$ cm$^{-2}$ Hz$^{-1}$\fi}
\newcommand\cc{\ifmmode {\rm~cm}$^{-3}$ \else cm$^{-3}$\fi}
\newcommand\cs{\ifmmode {\rm~cm}$^{-2}$ \else cm$^{-2}$\fi}
\newcommand\FWHM{\ifmmode {\rm~FWHM} \else ${\rm~FWHM}$\fi}
\newcommand\Msun{\ifmmode M_{\odot} \else $M_{\odot}$\fi}
\newcommand\Lsun{\ifmmode L_{\odot} \else $L_{\odot}$\fi}

\newcommand\hbeta{\ifmmode {\rm H}\beta \else H$\beta$\fi}
\newcommand\Kalpha{\ifmmode {\rm K}\alpha \else K$\alpha$\fi}
\newcommand\nh{\ifmmode N_{\rm H} \else N$_{\rm H}$\fi}

\title[Mrk~6: Long-term X-ray Study]{Long-term X-ray temporal and spectral study of a Seyfert galaxy Mrk~6}

\author[ N. Layek et al.]{
Narendranath Layek$^{1,2}$\thanks{E-mail: narendral@prl.res.in, narendranathlayek2017@gmail.com (NL)},
Prantik Nandi$^{1}$,
Sachindra Naik$^{1}$,
Neeraj Kumari$^{1}$,
Arghajit Jana$^{3}$, 
\newauthor
and Birendra Chhotaray$^{1,2}$
\\
$^{1}$Astronomy and Astrophysics Division, Physical Research Laboratory, Navrangpura, Ahmedabad 380009, Gujarat, India\\
$^{2}$Indian Institute of Technology, Palaj, Gandhinagar 382055, Gujarat, India\\
$^{3}$Instituto de Estudios Astrof\'isicos, Facultad de Ingenier\'ia y Ciencias, Universidad Diego Portales, Av. Ej\'ercito Libertador 441, Santiago, Chile \\
}

\date{Accepted XXX. Received YYY; in original form ZZZ}

\pubyear{2023}

\begin{document}
\label{firstpage}
\pagerange{\pageref{firstpage}--\pageref{lastpage}}
\maketitle

\begin{abstract}
We present a long-term X-ray study of a nearby Active Galactic Nucleus Mrk~6, utilizing observations from \xmm, \suz, \swift~ and \nustar~ observatories, spanning 22 years from 2001 to 2022. From timing analysis, we estimated variance, normalized variance, and fractional rms amplitude in different energy bands.The temporal study shows fractional rms amplitude ($F_{\rm var}$) below $10\%$ for the shorter timescale $(\sim60 ~\rm ks)$ and above $20\%$ for the longer timescale ($\sim \rm weeks$ ). A complex correlation is observed between the soft $(0.5-3.0$ keV) and hard $(3.0-10.0$ keV) X-ray bands of different epochs of observations. This result prompts a detailed investigation through spectral analysis, employing various phenomenological and physical models on the X-ray spectra. Our analysis reveals a heterogeneous structure of the obscuring material surrounding Mrk~6. A partially ionized absorber exhibits a rapid change in location and extends up to the narrow line regions or torus. In contrast, another component, located far from the central engine, remained relatively stable. During the observation period, the source luminosity in the 3.0--10.0 keV range varies between $(3-15) \times10^{42}$ \ergsec.

\end{abstract}


%%%%%%%%%%%%%%%%%%%%%%%%%%%%%%%%%%%%%%%%%%%%%%%%%%

%%%%%%%%%%%%%%%%% BODY OF PAPER %%%%%%%%%%%%%%%%%%

\section{Introduction}

 Active Galactic Nuclei (AGNs) are the extremely luminous and most persistent energetic sources in the universe. This extreme luminosity is powered by mass accretion onto the supermassive black hole (SMBH) residing at the centre of its host galaxy \citep{1984ARA&A..22..471R}. The AGNs emit in the entire band of the electromagnetic spectrum, starting from radio to gamma-rays. The X-ray emission from AGN is vital to probe the physical processes in extreme gravity as it is thought to originate from a high-temperature electron cloud called the corona, situated near the black hole \citep{1991ApJ...380L..51H,1994ApJ...428L..13N,1995ApJ...455..623C,2007A&ARv..15....1D}. The X-ray spectrum of an AGN is primarily characterized by the power-law continuum emission produced through the inverse-Comptonization \citep{ST1980} of the seed optical/UV photons from the standard accretion disc \citep{SS1973}. The primary power-law continuum gets reprocessed in the accretion disc and/or molecular torus and produces a reflection hump above 10 keV, an iron emission line at $\sim$ 6.4 keV \citep{1991MNRAS.249..352G,1991A&A...247...25M}, and soft excess emission below 2 keV \citep{1984ApJ...281...90H,1985A&AS...60..425A}. Depending on the presence or absence of broad optical emission lines, the AGNs are classified as Type~1 or Type~2. The classification of AGNs can be described using the simplified unification model based on the inclination angle of the obscuring torus \citep{1993ARA&A..31..473A}. In the optical/UV band, the `Type~1' AGNs show both broad ($\geq$1000\kms) and narrow ($\leq1000$ \kms) emission lines, while the `Type~2' sources show only narrow emission lines. Several studies use a finer classification scheme based on increasingly fainter broad emission lines (i.e., Type~1.2, 1.5, 1.8, and 1.9). Optical observations have identified a new subclass of AGNs called changing-look AGNs (CLAGNs). These objects display the appearance or disappearance of the broad optical emission lines, switching from Type~1 (or Type~1.2/1.5) to Type~2 (or Type~1.8/1.9) and vice versa in a timescale of months to decades \citep{2022arXiv221105132R}. These optical CLAGNs are also known as Changing-State AGNs. In X-rays, a different type of changing-look events are observed with the AGN switching between Compton-thin (line of sight hydrogen column density, $N_{H}$ $\leq10^{22}$ \cs) and Compton-thick ($N_{H}$  $\geq10^{22} $\cs) states \citep{2002ApJ...571..234R, 2003MNRAS.342..422M}, known as Changing-obscuration AGNs. Over the last decade, the number of such AGNs has grown up showing dramatic changes in spectral state and flux in optical as well as X-ray bands, e.g., UGC~4203 \citep{2010MNRAS.406L..20R}, NGC~4151 \citep{2007MNRAS.377..607P}, NGC~2992 \citep{1996ApJ...458..160W, 2007ApJ...666...96M}, IC~751 \citep{2016ApJ...820....5R}, NGC~6300 \citep{2002MNRAS.329L..13G, 2020MNRAS.499.5396J}. 

Markarian~6 (Mrk~6 or IC~450) is a nearby (z=0.0186\footnote{\url{https://ned.ipac.caltech.edu/byname?objname=MRk+6&hconst=67.8&omegam=0.308&omegav=0.692&wmap=4&corr_z=1}}) AGN that falls into the optical classification of an early-type S0 galaxy \citep{Osterbrock1976} with the central black hole mass of $M_{BH}\sim1.5\times10^8 M_{\odot}$ \citep{2014MNRAS.440..519A}. Considering its optical characteristics, Mrk~6 is commonly categorized as a Seyfert~1.5 AGN. However, it is noted that this source exhibits a ``changing-look'' behavior over time. Extensive studies of Mrk~6 have been conducted across various wavelengths, from radio to optical range, revealing the intricate nature of this AGN. From the optical observations, Mrk~6 was initially classified as an intermediate Seyfert galaxy \citep{Osterbrock1976}. Later, it was found that this source displayed Seyfert~1.5 characteristics in 1976 \citep{1983ApJ...265...92M}, underwent a transition to Seyfert~1.8 in 1977 \citep{2003ASPC..290...89D}, switched to a Seyfert~1.5 nature in 1979 \citep{1983ApJ...265...92M}, and consistently maintained its Seyfert 1.5 classification through the year 2010 \citep{2014MNRAS.440..519A}. Thus, from an optical perspective, Mrk~6 displays a changing-look \citep{2019sf2a.conf..509M, 2022ApJ...927..227L} behavior spanning the years from 1977 to 2010. It was also found that the optical line profiles of Mrk~6 exhibit noticeable variations over periods of months to years, indicating that some of the gaseous material responsible for these lines undergo coherent variations (\citealt{1992ApJS...81...59R}; \citealt{1993ApJ...409..584E}). The spectroscopic study showed the presence of broad Balmer lines and a strongly variable continuum in Mrk~6 \citep{1971Afz.....7..389K, 1993ApJ...409..584E, 2012MNRAS.426..416D}.

Radio observations have unveiled a complex structure surrounding the central AGN, characterized by a double set of bubbles and radio jets  \citep{Kukula1996}, suggesting a scenario of jet precession \citep{Kharb2006}. The structure of these jets remarkably resembles those observed in NGC~4151, leading to Mrk~6 often being referred to as a `4151 analog’ \citep{Capetti1995, Schurch2006} in both optical and X-ray studies. 

Although Mrk~6 has been well-studied in longer wavelengths, it was not extensively studied in the X-ray range (above 0.1 keV) until 1999. From the spectral analysis of a 40 ks ASCA observation, \cite{Feldmeier1999} first reported that Mrk~6 had a complex and heavy column density structure around the central engine. In this work, they interpreted this heavy absorption in terms of a partial covering model, thereby resolving the discrepancies in column density measurements along the line of sight using near-infrared/optical and X-ray data. Besides this, they also reported the presence of Fe \Kalpha~ line for the first time in this source, over the 0.5--10.0 keV X-ray primary continuum. Using  {\it BeppoSAX} observations, it was found that the density of the absorbing material is considerably variable on the time scale of two years \citep{Immler2003, 2003A&A...406..105M}. Further  \cite{2003A&A...406..105M} interpreted that the absorption originates in the broad line region (BLR). However, the nature of the absorbing medium (neutral, ionized, or a combination of both) remained unclear from their observations. Further, from the \xmm~ observation of Mrk~6 in 2003, \cite{Schurch2006} reported that the absorbing medium along the line of sight exhibited outflow characteristics, resembling a disc wind in nature.

In this work, we present our comprehensive findings of the long-term ($\sim$ 22 years; from 2001--2022) X-ray observations of Mrk~6 from various X-ray satellites. The paper is structured in the following way: Section~\ref{sec:OBS} provides an overview of the observational data and outlines the procedures used for data reduction. Detailed analysis of temporal behaviors and spectra of the source are presented in Section~\ref{sec:timing} and Section~\ref{sec:spec}, respectively. Then, we discuss our key findings in Section~\ref{sec:discussion}, and finally, our conclusions are summarized in Section~\ref{sec:conclusion}.

\section{OBSERVATIONS AND DATA REDUCTION}
\label{sec:OBS}

In this work, we utilize publicly available archival data for Mrk~6 obtained from \xmm, \nustar, \Swift, and \suz~observatories, accessed through HEASARC\footnote{\url{http://heasarc.gsfc.nasa.gov/}}. All data sets are reduced and analyzed using \texttt{HEAsoft} v6.30.1. 

\subsection{XMM-Newton}
Mrk~6 was observed with \xmm~\citep{2001A&A...365L...1J} at three epochs between March 2001 and October 2005. 
 We use the Science Analysis System ({\tt SAS v16.1.02}\footnote{\url{https://www.cosmos.esa.int/web/xmm-newton/sas-threads}} ) to reprocess the raw data from EPIC-pn \citep {2001A&A...365L..18S}. However, the EPIC-pn spectrum for the 2003 \xmm~ observation could not be obtained, as reported by \cite{2011ApJ...731...21M}. So, our analysis is restricted to the remaining \xmm~ (2001 \& 2005) observations.The details of the observations are presented in Table~\ref{tab:obs}. We consider only unflagged events with {\tt PATTERN $\leq 4$} in our analysis. We exclude the flaring events from the data by choosing appropriate {\tt GTI} files. The data are corrected for pile-up effect by considering an annular region with outer and inner radii of 30 arcsecs and 5 arcsecs, respectively, centred at the source coordinates while extracting the source events. We use a circular region of 60 arcsec radius, away from the source position, for the background products. The response files ({\it arf} and {\it rmf} files) for each EPIC-pn data set are generated by using the SAS tasks {\tt ARFGEN} and {\tt RMFGEN}, respectively. 

\subsection{NuSTAR}
\nustar~ is a hard X-ray focusing telescope consisting of two identical focal plane modules, FPMA and FPMB, and operates in the 3–79 keV energy range \citep{2013ApJ...770..103H}. Mrk~6 was observed with \nustar~simultaneously with \suz~ and \Swift~ in April 2015 and December 2015. The observation details are presented in Table~\ref{tab:obs}. We reprocess the data sets with the \nustar~ Data Analysis Software {\tt (NuSTARDAS v2.1.2\footnote{\url{https://heasarc.gsfc.nasa.gov/docs/nustar/analysis/}})} package. The standard {\tt NUPIPELINE} task with the latest calibration files CALDB \footnote{\url{http://heasarc.gsfc.nasa.gov/FTP/caldb/data/nustar/fpm/}} is used to generate the cleaned event files. The {\tt NUPRODUCTS} task is utilized to extract the source spectra and light curves. We consider circular regions of 60~arcsec and 120~arcsec radii for the source and background products, respectively. The source region is selected with the center at the source coordinates, and the background is chosen far away from the source to avoid any contamination.

\subsection{Swift}
The \textit{Swift} X-ray Telescope (XRT; \citealt{2005SSRv..120..165B}) is an X-ray-focusing telescope that operates in the energy range of 0.2--10.0 keV. Mrk~6 was monitored with the \Swift~ many times from 2006 to 2022. We stack these observations into six distinct instances labeled XRT1, XRT2, XRT3, XRT4, XRT5, and XRT6. Among these, four observations exhibit significantly longer exposure times than XRT1 and XRT2. As a result, a subsequent categorization is performed, isolating XRT3, XRT4, XRT5, and XRT6 for further analysis. To investigate the spectral variability, each of these four observations is divided into three segments: a, b, and c (see Table~\ref{tab:obs}). To extract spectra and light curves, we use the online tool `XRT product builder'\footnote{\url{http://swift.ac.uk/user_objects/}}\citep{2009MNRAS.397.1177E} provided by the UK Swift Science Data Centre. The tool processes and calibrates the data and produces final spectra and light curves of Mrk~6 in two modes, e.g., window timing (WT) and photon counting (PC) modes. 

\subsection{Suzaku}
{\it Suzaku} observed Mrk~6 on 21 April 2015 (Obs ID: 710001010) for an exposure of $\sim63$ ks with the X-ray imaging spectrometer (XIS) \citep{Koyama2007}. The photons were collected in the $3\times3$ and $5\times5$ editing modes. We use the standard data-reduction technique as described in the {\it Suzaku} Data Reduction Guide\footnote{\url{https://heasarc.gsfc.nasa.gov/docs/suzaku/analysis/abc/}}. We follow the recommended screening criteria while extracting {\it Suzaku}/XIS spectra and light curves. For this, we use the latest calibration files\footnote{\url{www.astro.isas.jaxa.jp/suzaku/caldb/}}, released on 3 February 2014, in {\tt FTOOL6.25} to reprocess the event files. The spectra and light curves for the source are extracted by considering a circular region of radius 200 arcsec centred at the coordinates of Mrk~6. For the background, we consider a circular region of 250 arcsec radius away from the source. The final spectra and light curves of Mrk~6 are generated by merging the data from the front-illuminated detectors (XIS0 and XIS3). The response files are generated using the task {\tt XISRESP}. It is important to note that we ignore the known Si~K edge in the spectrum by avoiding the data from 1.6 keV to 2.0 keV.

\begin{table}

	\centering
	\caption{Log of observations of Mrk~6}
	\label{tab:obs}
	\begin{tabular}{l c c c r} 
	\hline
    ID  &   Date     &         Obs. ID    &     $\rm Observatory^{\dagger}$    &     Exposure\\
   &  (yyyy-mm-dd)   &                          &                    &   (ks)     \\
    \hline
 XMM1 & 2001-03-27   &  0061540101   &    \xmm    &    46.5  \\
                                         \\
XMM2 &  2005-10-27     &  0305600501     &  \xmm   & 21.8  \\  
                                          \\
XRT1 &  2006-01-19    & 00035461001      &    \swift    &  14.9  \\
     &                & -00035461004    &                  &    \\
                                \\
SU &  2015-04-21   &  710001010      &    Suzaku    &       63.0 \\
                           \\
NU1  &  2015-04-21   & 60102044002     &     \nustar        &      62.5\\
               \\

XRT2 &  2015-11-08  &   00081698001     &   \swift        &    19.1   \\
     &              &   -00081698003    &                &            \\
NU2 &   2015-11-09   &  60102044004     &   \nustar   &       43.8  \\
                    
                                       \\
               
XRT3a &  2019-02-27   &  00035461005     &   \swift    &      6.5  \\
      & - 2019-09-28  &  -00035461011    &                  &             \\
XRT3b & 2019-11-9     &   00035461012    &    \swift  &        6.1     \\
      & -2019-11-29   &   -00035461016   &                  &                \\
XRT3c &  2019-12-05   &   00035461017    &    \swift    &       6.8       \\
      &  -2019-12-25  &   -00035461020   &                  &                  \\
                                  \\
XRT4a &  2020-01-02  &   00035461021   &     \swift   &    14.1      \\
      &  -2020-03-26 &   -00035461032  &                  &               \\
XRT4b &  2020-04-10  &  00035461033    &    \swift   &         14.8     \\
      & -2020-07-29  &  -00035461043   &                  &                   \\
XRT4c &  2020-08-2   &  00035461044    &    \swift    &        17.1     \\  
      &  2020-12-25  &  -00035461052   &                  &                 \\
                                   \\
XRT5a &  2021-01-08   &  00035461053     &   \swift   &    19.6    \\
      &  -2021-04-17  &  -00035461061   &                 &            \\
XRT5b &  2021-05-01   & 00035461062  &   \swift       &    18.6       \\       
      &  -2021-08-15  & -00035461070  &                   &                \\
XRT5c & 2021-09-12    & 00035461071 &   \swift       &    25.4        \\      
      & -2021-12-30   &  -00035461080 &                   &                   \\
                    \\
XRT6a &  2022-01-13   &  00035461081    &    \swift    &    16.3   \\
      &  -2022-04-21   & -00035461190   &                  &           \\
XRT6b &  2022-05-04   &  00035461091    &    \swift   &    18.5        \\
      &  -2022-08-11  &  -00035461098   &                  &                \\
XRT6c &  2022-09-08   &  00035461099    &    \swift   &  19.4       \\
      & -2022-12-29   &  -00035461107   &                  &                \\

		\hline
	\end{tabular}
\leftline{$\dagger$ Data from the \xmm/EPIC-PN, \swift/XRT, FPMA \& FPMB of }
 \leftline{\nustar, and \suz/XIS instruments are used in this work.}
\end{table}
\section{DATA ANALYSIS AND RESULTS}
\label{sec:Data Analysis and Results}

\subsection{Timing analysis}
\label{sec:timing}
\begin{figure*} 
	\centering
	\includegraphics[scale=1.2]{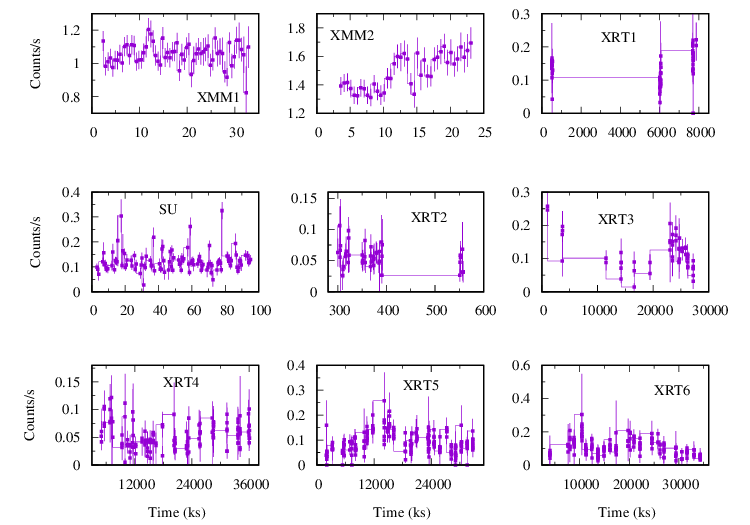}
	\caption{Variation of photon counts with respect to time from \xmm, \suz~ and {\it Swift}/XRT observations of Mrk~6 at different epochs (see Table~\ref{tab:obs}). The light curves in the $0.5-10.0$ keV range are shown here.}
	\label{fig:lightcurves} 
\end{figure*}

 \begin{table*}
	\centering
	\caption{The table provides variability statistics in different energy ranges for various observations, calculated using light curves with 500s time bin. It is important to mention that in most of the cases, the average error of the observational data surpasses the 1$\sigma$ limit, resulting in negative excess variance. As a result, these cases contain imaginary values for $F_{\rm var}$, and thus, they are excluded from the table.}
	\label{tab:Fvar}
	\begin{tabular}{l c c c c c c c c c} 
	\hline
ID &  Energy band & $N$ & $x_{\rm max}$ &  $x_{\rm min}$ & $\mu$ & $R=\frac{x_{\rm max}}{x_{\rm min}}$ & $\sigma^{2}_{\rm NXS}$ &  $F_{\rm var}$ \\

    &  keV        &     & count $s^{-1}$ & count $s^{-1}$ & count $s^{-1}$    &    &      &$\%$ \\
                       \hline
    XMM1 & 0.5-3.0  &  64  & 1.640   &0.327   & 0.450& 5.00   & 0.021 $\pm$0.020     &14.0$\pm$ 7.1\\
         & 3.0-10.0  & 64  & 1.142  & 0.452  & 0.633 & 2.52  & -0.010      & --            \\
         & 0.5-10.0  & 63  & 1.204  & 0.824  & 1.053  & 1.46& -0.002 &-- \\
         \hline
    XMM2 & 0.5-3.0  & 40   & 0.757   & 0.491   &0.619  & 1.54 & 0.003 $\pm$0.002 & 5.6$\pm$ 2.0   \\ 
         & 3.0-10.0  & 40   &1.036    & 0.740   &0.871 &1.39   & 0.003$\pm$ 0.016 & 5.3$\pm$1.6    \\ 
         &0.5-10.0   & 40   & 1.694   & 1.311   & 1.490 &1.29  & 0.003$\pm$0.001  & 5.8$\pm$1.2\\
                    \hline
    NU1 & 3.0-6.0  &  151  &  0.103   & 0.017  & 0.065  &6.50 & -0.006   &  --    \\ 
        & 7.0-60.0 &  156  & 0.603   & 0.147  & 0.261 &4.10  &-0.004    & --\\
        & 3.0-60.0   & 157  & 0.643    &0.246     & 0.361 & 2.45 & -0.008 & --\\
                        \hline 
    NU2 & 3.0-6.0  & 110   &0.296 & 0.076   & 0.134  & 3.86 &-0.003     & --     \\ 
        & 7.0-60.0 &  111  & 0.920 & 0.218  &0.405    &4.21  &0.008$\pm$0.004& 9.3$\pm$ 0.2 \\
        & 3.0-60.0   & 112  & 5.244    &0.437     & 0.635 & 12.00 & -0.054 & --\\
        \hline
    XRT1 & 0.5-3.0 &52   &0.089   & 0.007   & 0.047   &12.00   &0.053$\pm$0.040  &23.0$\pm$9.0   \\  
         & 3.0-10.0 &53   &0.154   &0.014    &0.086    &11.00   &0.049$\pm$0.026  &22.0$\pm$6.3   \\
         & 0.5-10.0 & 53  &0.221   &0.032    &0.132    &6.87   &0.037$\pm$0.020   &19.0$\pm$5.7   \\
       \hline
     XRT2 & 0.5-3.0 &40   &0.041   &0.002    &0.010    &18.28   &-0.107  & --  \\  
         & 3.0-10.0 & 48  &0.091   &0.013    &0.047    &6.63   &-0.016  &--  \\
         & 0.5-10.0 & 47  &0.106   &0.025    &0.055    &4.08   &-0.008   &--   \\
\hline
    XRT3 & 0.5-3.0 & 52  &0.114   & 0.002   & 0.050   & 49.28  &0.168$\pm$0.046  &4.1$\pm$6.9   \\  
         & 3.0-10.0 &54   &0.152   &0.014    &0.067   &10.35   &0.087$\pm$0.031 &29.0$\pm$5.9   \\
         & 0.5-10.0 &53   &0.257   &0.031    &0.117    &8.08   &0.101$\pm$0.026   &31.0$\pm$5.0   \\
\hline
    XRT4 & 0.5-3.0 &130  &0.071   &0.002    &0.023    &33.26   &-0.006  &--   \\  
         & 3.0-10.0 &128   &0.083   &0.005    &0.031    &15.72   &0.075$\pm$0.030  &27.0$\pm$5.7   \\
         & 0.5-10.0 &125   &0.122  &0.015    &0.055    &7.78   &0.022$\pm$0.020   &15.0$\pm$6.9   \\
\hline

XRT5 & 0.5-3.0 & 179  &0.099   &0.004    &0.034    &20.39   &0.017$\pm$0.022  &13.0$\pm$8.7   \\  
         & 3.0-10.0 &185   &0.180   &0.012    &0.061    &14.36   &0.117$\pm$0.021  &34.0$\pm$3.6   \\
         & 0.5-10.0 & 185  &0.257   &0.020    &0.093    &12.40   &0.080$\pm$0.016  &28.0$\pm$3.3   \\

      \hline   
XRT6 & 0.5-3.0 &156   &0.202   &0.004    &0.046    &42.27   &0.051$\pm$0.034  &22.0$\pm$7.7   \\  
         & 3.0-10.0 &159  &0.151   &0.009    &0.064    & 16.47  &0.082$\pm$0.023  &28.0$\pm$4.3   \\
         & 0.5-10.0 &161   &0.303   &0.030    &0.107    &10.01   &0.073$\pm$0.020   &27.0$\pm$4.1   \\
\hline
 XRT$^\dagger$ & 0.5-3.0 &511   &0.098   &0.001    &0.011    &55.04   &-0.209  &--   \\  
         & 3.0-10.0 &637   &0.293   &0.008    &0.083   &35.80 &0.196$\pm$0.013   &44.0$\pm$1.9     \\
         & 0.5-10.0 &626   &0.391   &0.021    &0.093    &18.07   &0.146$\pm$0.012   &38.3$\pm$2.0  \\
         \hline
        
	\end{tabular}
\leftline{$\dagger$ Represent combined $\Swift$ observation from XRT1(MJD-53754)-XRT6(MJD-59886).}
\end{table*}

The timing analysis is carried out on the X-ray light curves obtained from the \xmm, \nustar, and \Swift~ observations of Mrk~6 (see Table~\ref{tab:obs}). The time resolution of the light curves used in our analysis is set at 500 s. The light curves in the 0.5-10 keV range, generated from the \xmm~ and ~\Swift~ observations, are shown in Figure~\ref{fig:lightcurves}. Additionally, we utilize the light curves of combined \Swift~ observations, employing a bin size of one day for our correlation study. Further, we extract light curves from the \xmm~ and \Swift~ observations in two energy bands, the soft X-ray band in the 0.5–3.0 keV range and the hard X-ray band in the 3.0–10.0 keV range for variability analysis. Light curves from the NuSTAR observations in the 3.0–60.0 keV energy range are utilized to explore the variability in the high-energy regime. The light curves in the entire energy range (3.0–60.0 keV) are divided into two energy bands, such as band1 (3.0–6.0 keV range ) and band2 (7.0–60.0 keV range). We carefully avoid the 6.0 – 7.0 keV band due to the presence of the Fe~\Kalpha~ line in this range.

\subsubsection{Fractional variability}
To check the temporal variability of Mrk~6 across different energy bands, we calculate the fractional variability $F_{\rm var}$ (\citealt{1996ApJ...470..364E}; \citealt{1997ApJ...476...70N};   \citealt{1997ApJS..110....9R}; \citealt{2003MNRAS.345.1271V}; \citealt{2012ApJ...751...52E}). The fractional variability for a light curve of $x_i$ counts/s with the measurement error $\sigma_{i}$ for $N$ number of data points, mean count rate $\mu$ and standard deviation $\sigma$, is given by the relation,
\begin{equation}
F_{\rm var}=\sqrt{\frac{{\sigma^2_{\rm XS}}}{\mu^2}} 
\end{equation}
where, $\sigma^2_{\rm XS}$ is the excess variance (\citealt{1997ApJ...476...70N}; \citealt{2002ApJ...568..610E}), used to estimate the intrinsic source variance and given by,
\begin{equation}
\sigma^2_{\rm XS}=\sigma^2- \frac{1}{N} \sum_{i=1}^N \sigma^2_{i}
\end{equation}
Normalized excess variance is defined as $\sigma^2_{\rm NXS}=\sigma^2_{\rm XS}/\mu^2$. The uncertainties in $\sigma^2_{\rm NXS}$ and $F_{\rm var}$ are estimated as described in \cite{2003MNRAS.345.1271V} and \cite{2012ApJ...751...52E}. The peak-to-peak amplitude is defined as $R = x_{\rm max} /x_{\rm min}$ (where, $x_{\rm max}$ and $x_{\rm min}$ are the maximum and minimum count rates, respectively) to investigate the variability in the X-ray light curves.

\begin{figure*}
     \centering
     \hspace{-0.2in}
     \begin{subfigure}[b]{0.25\textwidth}
         \includegraphics[width=1.7\textwidth]{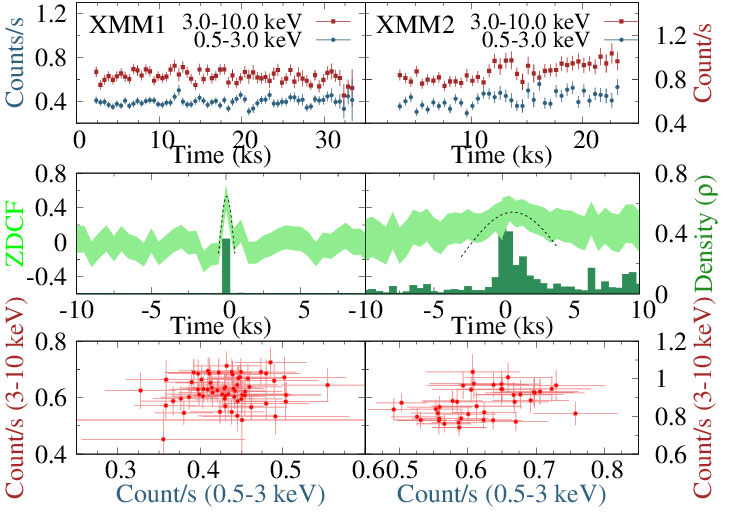}
         \caption{\xmm}
     \end{subfigure}
     \hfill
     \hspace{0.9in}
     \begin{subfigure}[b]{0.25\textwidth}
         \centering
         \includegraphics[width=1.7\textwidth]{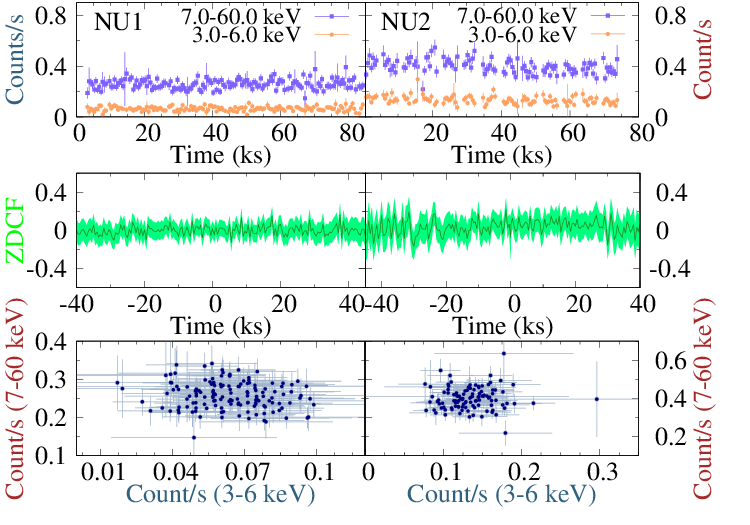}
         \caption{\nustar}
         \end{subfigure}
     \hfill
     \hspace{0.4in}
     \begin{subfigure}[b]{0.25\textwidth}
         \centering
         \includegraphics[width=1.7\textwidth]{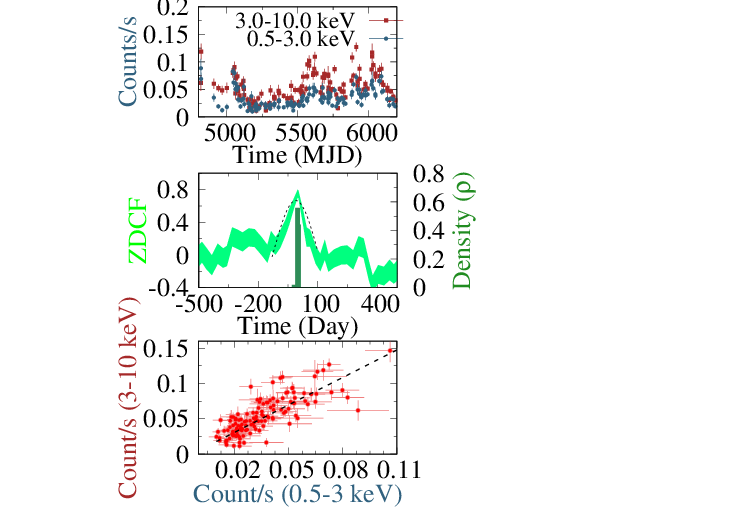}
         \caption{\Swift}
         \end{subfigure}
        \caption{The multi-plots display the light curves in different energy ranges, the correlation between the light curves, and the count-count plot from three different X-ray instruments (\xmm,~\nustar,~and \Swift). {\it Top left panels:} The light curves of Mrk~6 in 0.5 to 3.0 keV (blue) and 3.0 to 10.0 keV (brown) ranges are presented for two epochs of \xmm~(XMM1 and XMM2) observations. The top middle panels show the light curves of the source in 3.0 to 6.0 keV (blue) and 7.0 to 60 keV (brown) ranges for two epochs of \nustar~ (NU1 and NU2) observations. The top right panel shows the light curves of the source in 0.5 to 3.0 keV (blue) and 3.0 to 10 keV (brown) ranges for all the \Swift~ observations of Mrk~6 between 2019 and 2022. It can be seen that the average count rate of the source in the high energy bands is high in comparison to the low energy band for all the observations. \\ {\it Middle panels: } Corresponding ZDCF (light-green) analysis curves showing the correlation as a function of time delay between the X-ray light curves are plotted. The likelihood functions (dark green), simulated using 102000 points, are plotted along with the ZDCF.\\ {\it Lower panels:} The count vs. count plots are presented for all the observations.}
        \label{fig:correlation}
\end{figure*}

For both the \xmm~ observations of Mrk~6 (XMM1 \& XMM2), the results of the variability analysis for different energy bands are given in Table~$\ref{tab:Fvar}$. During the XMM1 observation, we find that the peak-to-peak amplitude $R$ varies in the range of 5.00 to 1.46. However, due to the low count rate and large error associated with each data point, we encounter negative values for $\sigma^2_{\rm NXS}$, resulting in imaginary fractional variability ($F_{\rm var}$) for light curves in hard X-ray band and entire energy band. In the soft X-ray  band , $\sigma^2_{\rm NXS}$ and $F_{\rm var}$ are estimated to be 0.021$\pm$0.020 and 14$\%$, respectively. Similarly, during the XMM2 observation, we find that $R$ varies in the range of 1.54 to 1.29. However, $\sigma^2_{\rm NXS}$, and $F_{\rm var}$ are approximately the same for the soft, hard, and the entire energy bands with values of $0.003\pm0.002$ and $\sim5.6\%$, respectively. For \nustar~ observations (NU1 \& NU2), the results of the variability analysis ($x_{\rm max}$, $x_{\rm min}$, $\mu$) are presented in Table $\ref{tab:Fvar}$. The variation of the X-ray photons with time in band1 and band2 for both the observations (NU1 \& NU2) are shown in the top middle panels of Figure~\ref{fig:correlation}.

For \Swift~ observations, the variability parameters like $\sigma^2_{\rm NXS}$ and corresponding ($F_{\rm var}$ ) are also calculated. However, due to the low count rate and high error associated with each data point, we encounter negative values for normalized excess variance, resulting in imaginary fractional variability for the XRT2 observation. For observations with positive $\sigma^2_{\rm NXS}$, the observed $F_{\rm var}$ is found to be ranging from 4.1$\%$ to 23$\%$ with an average of 16$\%$ for the soft X-ray band. However, we find negative normalized excess variance for XRT4 in this energy band. In the hard X-ray band, $F_{\rm var}$ is found to be 22$\%$ to 34$\%$ with an average of 28$\%$. For the entire energy band, we obtain $F_{\rm var}$ ranging from 15$\%$ to 31$\%$ with a mean of $24\%$. We also calculate the variability parameters of the combined $\Swift$ observations (XRT1--XRT6). In this case, we obtain $F_{\rm var}$ as 44$\%$ for the hard X-ray band and 38\% for the entire energy band. However, the average count rate in the soft X-ray band is very low (0.001 count/s), causing noise dominance and resulting in negative normalized excess variance. The details of variability analysis are presented in Table~\ref{tab:Fvar}.

Temporal variability in different energy bands provides insight into the physical properties of the emitting region. In the case of Mrk~6, the normalized excess variances were close to zero in the \xmm~( XMM1--MJD 51995) and ~\nustar~( NU1 \& NU2) observations, indicating the uncertainties of observed data surpassing the data dispersion. This implies insignificant variability above the count rate uncertainties, except for XMM2 (MJD--53670), which exhibited only $\sim5.7\pm1.5 \%$. In the combined \Swift~ observations, over $20\%$ variability was detected. So, we can conclude that Mrk~6 shows temporal variability below $10\%$ in shorter timescales ($\sim$60  ks), whereas, for a longer timescale ($\sim$ \rm weeks), we observe over $20\%$ temporal variability.

\subsubsection{Correlation}
\label{Correlation}

To investigate the correlation between the light curves, we conduct a detailed cross-correlation analysis of short-term and long-term X-ray observations of Mrk~6. We use two epochs of \xmm~(XMM1 \& XMM2) and \nustar~ (NU1 \& NU2) observations for the short-term study, whereas we employ the combined \Swift~observations (XRT3--XRT6) for the long-term analysis.

We use the $\zeta$-transformed discrete correlation function ($\zeta-$DCF\footnote{\url{www.weizmann.ac.il/particle/tal/research-activities/software}}, \citealt{1997ASSL..218..163A}) method to investigate the correlation between the variation of photon counts in different energy bands. To determine the significance of the correlation function, we utilize the likelihood function for each discrete correlation function (DCF). For this purpose, we use 102000 simulated points in the {\tt $\zeta-$DCF} code for the light curves obtained from the \xmm, \nustar, and \Swift~ observations. The error in the position of the peaks is calculated using the formula given in \cite{1987ApJS...65....1G}, and the corresponding values are given in Table~\ref{tab:ZDCF}. The light curves with different energy bands from \xmm, \nustar, and combined \Swift~ observations are shown in the left, middle, and right top panels of Figure~\ref{fig:correlation}, respectively. Furthermore, we plot the correlation function with corresponding light curves in the middle panels of Figure~\ref{fig:correlation}. The count-count plots are also presented in the bottom panels of the same figure.
 
We begin our analysis using the 2001 \xmm~ observation (XMM1, MJD--51995). We find that the bin size of the light curve is larger than the value of uncertainty calculated using the formula given by \cite{1987ApJS...65....1G}. So, we consider the bin size as the uncertainty on the position of the peak. We use a similar approach to estimate the delay in the light curves for the XMM2 observation. The estimated values are presented in Table~\ref{tab:ZDCF}.

In the case of \nustar~ observations,  it is found that these two bands (band1 \& band2 ) are uncorrelated. From the spectral analysis of these observations, we notice that below 10.0 keV, the X-ray photons originate from the Compton cloud, while above 10.0 keV, the reflection component dominates (see Section~\ref{sec:spec}). We do not find any correlation between band1 and band2 as we examine the correlation between two types of photons with distinct origins. 

\begin{table}
	\centering
	\caption{Details of the parameters utilized for estimating time delays between light curves in two different energy ranges. The peak value of the correlation function is denoted by $\epsilon^{z}_{\tau}$, and the corresponding time delay is represented by $\tau^{zdcf}_{d}$. Using the method outlined in \citep{1987ApJS...65....1G}, the error on the position of the peak of the correlation function is determined and is given by $\Delta\tau_{d}$. A comparison between this error and the time bin size was conducted to ensure the accuracy of the result, with the larger value selected for our analysis. For further details, refer to Section~\ref{Correlation} }
	\label{tab:ZDCF}
	\begin{tabular}{l c c c c c} % four columns, alignment for each
     \hline
     ID  &  Epochs & Bin size  &  $\Delta\tau_{d}$  & $\epsilon^{z}_{\tau}$  & $\tau^{zdcf}_{d}$   \\
         &   Year & (ks)  &  (ks)  &   & (ks) \\
                                     \hline
    XMM1 & 2001 & 0.5 &0.21    & $0.48\pm0.05$    &0.31$\pm$0.5  \\
    XMM2 & 2005 & 0.5 &0.46   & $0.50\pm0.03$    &0.77$\pm$0.5   \\
    NU1 & 2015 & 0.5 &  -  &  -   & - \\
    NU2 & 2015 & 0.5 &  -  &  -   &  - \\
    XRT$^{\dagger}$ & 2019-2022 & $1.0$   & $5.07$    &$0.68\pm0.05$      &$-1.88\pm5.07$     \\
    
    \hline
	\end{tabular}
 \leftline{$\dagger$ In the case of XRT observation, the correlation parameters are }
 \leftline { calculated in the unit of days.}
\end{table}

We then proceed to investigate the light curves from \Swift~ observations. Due to the limited resolution and low exposure time, we opt for a one-day bin size for the light curves from all XRT observations. The high uncertainty associated with smaller bin sizes, such as 500 seconds, led to the exclusion of many data points. We attempted different bin sizes for the light curves and found that the one-day bin is the optimum size for the temporal analysis. 
We consider 2019 to 2022 \Swift~ observations, representing nearly continuous observations for our analysis. We exclude XRT1 (2006) and XRT2 (2015) observations due to their significant temporal separation from the nearly continuous XRT observations during 2019–2022. The light curves for two different bands are shown in the top right panel of Figure~\ref{fig:correlation}. We use similar techniques to calculate the correlation function ($\zeta-$DCF). However, our analysis did not reveal any plausible time delay ($-1.88\pm5.07$ days) between the soft X-ray and hard X-ray bands. The correlation function has a peak value of $\epsilon^{z}_{\tau}=0.68\pm0.05$. The detection of correlation between soft X-ray (0.5--3.0 keV) and hard X-ray (3.0--10.0 keV) bands suggests that the photons in both energy bands could have originated through the same physical mechanism \citep{Neeraj2021, Nandi2021, Nandi2023}. On the other hand, we find that the \Swift~ spectra from 2019--2022 observations are well-fitted by a single absorption coefficient (see Section~\ref{sec:spec}). This finding indicates that the low-energy photons (below 3.0 keV) are less affected by column densities compared to other observations, where two absorption coefficients are required to fit the low-energy portions of each spectrum.

From the above study, it is evident that the light curves of soft X-ray and hard X-ray bands do not exhibit any significant correlation, or at most show very weak correlation during the \xmm~ observations in 2001 (XMM1) and 2005 (XMM2). These observations are further characterized by the need for two absorption coefficients to fit the spectra below 3.0 keV, indicating a complex structure in the low-energy spectrum. 
 
However, during the 2019–2022 \Swift~ observations, a reasonably strong correlation is observed between these energy bands. From the spectral study of data from these observations, we find that the low-energy part of the X-ray spectrum is relatively simple for Swift/XRT observations. It is well-fitted with a single absorption coefficient model. In the case of a high energy band (above 10.0 keV), we observe the domination of the reflection component over the primary continuum. As a result, the light curves in  band1 and band2 are found to be uncorrelated. The detailed results are given in Table~\ref{tab:ZDCF}, and the corresponding correlation functions are plotted in the middle panels of Figure~\ref{fig:correlation}.

\subsection{Spectral analysis}
\label{sec:spec}

We use data from \xmm, \nustar, \suz, and \Swift~ observations of Mrk~6 in our spectral analysis to investigate the spectral variations of the source over an extensive time frame of $\sim 22$ years (2001-2022). We use $\texttt {XSPEC}$ v12.12.1 \citep{1996ASPC..101...17A} software package for spectral analysis. The $\chi^2$ statistics is used to determine the best-fitting models to describe the observed data. The spectral analysis is carried out using \xmm~ observations in 2001 and 2005 in the 0.5-10 keV range, simultaneous \suz~ and \nustar~ observations in the 0.5--60 keV range, simultaneous \Swift~ and \nustar~ observations in the 0.5--60 keV range, and XRT observations for 13 epochs from 2006 to 2022 in the 0.5--10 keV range (see Table~\ref{tab:obs}). {\it NuSTAR} data beyond 60 keV are not considered in the present analysis as it is dominated by background. To ensure robust statistics, we bin the data in such a way that there are  at least 30 counts/bin for both \xmm, \suz~ and \nustar~ observations, while we used 10 counts in each bin for the \Swift~ observations. The {\tt GRPPHA} task is used for binning the spectral data. The quoted errors for best-fitting spectral parameters are determined at a 90 percent confidence level by using the command {\tt error} in {\tt Xspec}. We calculate the unabsorbed X-ray luminosity from each spectrum using {\tt clumin}\footnote{\url{https://heasarc.gsfc.nasa.gov/xanadu/xspec/manual/node286.html}} on the  {\tt powerlaw} model. We estimate the Eddington luminosity $L_{\rm Edd}=1.95\times10^{46}$ erg/s by using the relation, $ L_{\rm Edd}=1.3\times10^{38} (M_{\rm BH}/M_\odot)$, where $M_{\rm BH}=1.5\times10^8~M_\odot$ \citep{2014MNRAS.440..519A}. The bolometric luminosity is calculated using the intrinsic luminosity of the source in the energy range from 2.0 to 10.0 keV with the bolometric correction factor 20 \citep{Vasudevan2009}. While estimating the luminosity, we use the redshift, $z$=0.0186. Using the bolometric luminosity and Eddington luminosity of the source, we derive the Eddington ratio ($ \lambda_{\rm  Edd}$), which is defined as the ratio between the bolometric luminosity $ (L_{\rm bol})$ and Eddington luminosity $ (L_{\rm Edd})$. Throughout this work, we use the Cosmological parameters as follows: \( {H}_0 = 70 \, \text{km s}^{-1} \, \text{Mpc}^{-1}, \Lambda_0 = 0.73, \text{ and } \sigma_M = 0.27\) \citep{2003ApJS..148....1B}. 

\subsubsection{Model Construction}
\label{model const}
To investigate the spectral variability in the source during our observation period, we construct a base model covering the broad energy range from 0.5 to 60.0 keV. We begin by employing simple models, such as the power law, to characterize the observed X-ray spectra. Later, these models are replaced with more sophisticated phenomenological and physical models to understand the accretion dynamics and other physical properties. A  {\tt constant} component is used as a cross-normalization factor while using data from different instruments in simultaneous (SU+NU1 and XRT2+NU2) spectral fitting.

Initially, we consider the 3.0 to 10.0 keV X-ray continuum spectrum of the source for spectral fitting. According to current understanding, the X-ray continuum photons are produced through the process of inverse Compton scattering, wherein the thermal photons from the accretion disc are up-scattered in a hot electron cloud. This process can produce a power-law type of spectrum. Therefore, we consider {\tt powerlaw} model to fit the spectrum of each observation given in Table~\ref{tab:obs}. Along with this, we also consider Galactic hydrogen column density ($N_{\rm H,gal} $) along the line of sight as the multiplicative model {\tt TBabs} \citep{2000ApJ...542..914W} within {\tt XSPEC}. We fix the value of $ N_{\rm H,gal}$ at $7.63\times10^{20}$ \cs, the Galactic value in the direction of the source \footnote{\url{https://heasarc.gsfc.nasa.gov/cgi-bin/Tools/w3nh/w3nh.pl}}. Thus, our base model for the 3.0 to 10.0 keV X-ray spectral fitting is as follows: 

\begin{center}
    ${\tt TBabs \times const \times powerlaw}$
\end{center}

\begin{figure} 
	\centering
	\includegraphics[scale=0.7, width=8.7cm]{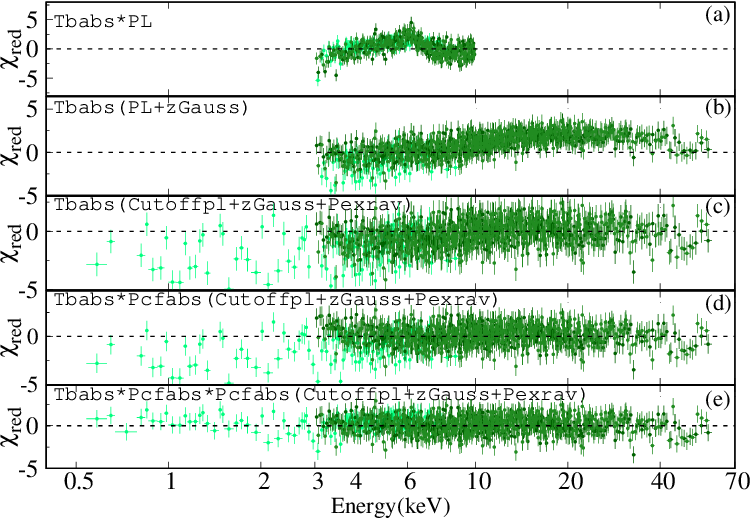}
	\caption{Variation of $\chi_{\text{red}}$ values is shown for each model across the broadband (XRT2+NU2) spectra of Mrk~6. The analysis started in the 3.0-10.0 keV range spectrum with the simple {\tt powerlaw} model, and corresponding residuals are shown in the top panel (a). Then, we added {\tt zGaussian } for the Fe-line at $\sim$ 6.45 keV. Later, we extended spectra above 10 keV. The corresponding residuals are shown in the second panel (b). The {\tt pexrav} model was used to account for the reflection component. Next, we extended the spectra below 3.0 keV, and the corresponding residuals are shown in the third panel (c). The residuals obtained by adding single and double {\tt pcfabs} components to account for the absorption characteristics of the local absorber in the soft X-ray ranges are shown in the fourth (d) and fifth (e) panels, respectively.}
	\label{fig:chi} 
\end{figure}

 We fit each spectrum with this model, and the corresponding variation of $\chi_{\rm red}$ for XRT2+NU2 (MJD--57335) observation is illustrated in the panel~(a) of Figure~\ref{fig:chi}. After the continuum fitting, we observe positive residuals in the $6-7$ keV range (see Figure~\ref{fig:chi}(a)), which is attributed to the presence of Fe~K$_\alpha$ line. To account for this, we introduced a Gaussian model, {\tt zGauss} in {\tt Xspec}. As a result, our model to fit the spectrum in the 3.0 to 10.0 keV range is as follows:

  \begin{center}
    ${\tt TBabs \tt \times const(powerlaw+zGauss)}$
\end{center}
 
 After successfully fitting the primary continuum and the iron K$_\alpha$ line in 3.0 to 10.0 keV  range (see Figure~\ref{fig:chi}(b)), we proceed to extend the X-ray spectra into the high energy regime (above 10 keV). In doing this, we find that the observational broadband data do not align with our model (Figure~\ref{fig:chi}(b)). We use another {\tt powerlaw} to address this deviation in the high-energy data points from the primary model. Subsequently, this additional power-law component is later substituted with {\tt pexrav}, and the corresponding variation of $\chi_{\rm red}$ is shown in Figure~\ref{fig:chi}(c). It is to be noted that the {\tt powerlaw} component is replaced by a cut-off power law ({\tt cutoffpl}) to investigate the presence of a high-energy cut-off in the broadband observations. So, for broadband observations, the  model became:
 \begin{center}
    ${\tt TBabs\times const(cutoffpl+zGauss+pexrav)}$
\end{center}

To address the low-energy counterpart of the observed X-ray spectra (below 3.0 keV), we initially employ a single {\tt pcfabs}\footnote{\url{https://heasarc.gsfc.nasa.gov/xanadu/xspec/manual/XSmodelpcfabs.html}} along with the ${\tt TBabs \times const(cutoffpl+zGauss+pexrav)}$ model. The corresponding variation of $\chi_{\rm red}$ is shown in Figure~\ref{fig:chi}(d) for the observation XRT2+NU2. It is evident in this panel that a single {\tt pcfabs} is insufficient to account for the local hydrogen column density along the line of sight. Therefore, we introduce another {\tt pcfabs} to fit the broadband spectrum in the 0.5--60.0 keV range, and the corresponding variation of $\chi_{\rm red}$ is shown in Figure~\ref{fig:chi}(e). As a result, the composite model employed to fit the broadband spectra can now be defined as follows:

\begin{center}
$ \tt { TBabs\times pcfabs\times pcfabs\times const(cutoffpl+zGauss+pexrav)}.$
\end{center}

To investigate the ionization properties of the absorber, we replace the second {\tt pcfabs} component with {\tt zxipcf} \citep{Miller2006, Reeves2008} component. It is worth mentioning that a previous study by \citet{Feldmeier1999}, \citet{Immler2003}, and  \citet{2003A&A...406..105M} also employed double absorption components and suggested that one absorber might be located in proximity to the black hole or within the BLR region, while the second absorber could be situated further away, possibly within the torus. This model ({\tt zxipcf}) can calculate the amount of ionization of the medium through the ionization parameter $\rm (\xi)$ as $\rm \xi=L/(nR^2)$, where $\rm L$ is the luminosity of the irradiating source, $n$ is the number density of the irradiated material and $ R$ is the distance between the source and the irradiated material.
 
 It is noted that while fitting data in the 0.5--10 keV range (XMMs and XRT observations in Table~\ref{tab:obs}), we use ${\tt TBabs \times pcfabs \times pcfabs \times (powerlaw+zGauss)}$ and ${\tt TBabs\times pcfabs \times zxipcf \times (powerlaw+zGauss)}$ as our composite models. 
 
\subsubsection{\tt powerlaw}
\label{sec:pl}

We started our spectral analysis with an absorbed power-law model with a Gaussian line ({\tt zGauss}) as described in section~\ref{model const}. The model in {\tt Xspec} reads as: ${\tt TBabs\times pcfabs\times pcfabs\times(powerlaw+zGauss)}$. This composite model fits the X-ray spectra up to 10.0 keV. Corresponding power-law indices are found to be $\Gamma=1.53\pm0.10$ and 1.57$\pm$0.10 for XMM1 and XMM2 observations, respectively. Next, we analyze the data obtained from the \Swift~ observations of the source for the energy range of 0.5--10 keV. Across all 13 XRT spectra, the Fe $\Kalpha$ line remains undetected. This could be due to the low exposure time of each observation, combined with the poor energy resolution of {\it Swift}/XRT. Therefore, we fit the $\Swift$ spectra by removing the \texttt{Gaussian} component from the baseline model. The resultant fitted model in {\tt Xspec} reads as: ${\tt TBabs \times pcfabs \times pcfabs \times powerlaw}$. From the analysis of XRT1 (MJD--53754), we find that the column densities increased and photon index ($\Gamma$) changed from, $\Gamma\sim$$1.57$ to $1.73$ after the XMM2 observation ( MJD--53670).

\begingroup
\begin{table*}
	\centering
	\caption{Parameters obtained from the ${\tt TBabs\times pcfabs\times pcfabs\times(powerlaw+zGauss)}$ model fitting with all the spectra  for the energy range of 0.5 to 10 keV}. The unabsorbed X-ray continuum luminosity is calculated for the energy range of 3.0 to 10.0 keV. The detailed results are discussed in Section~\ref{sec:pl}
	\label{tab:powerlaw}
 \setlength{\tabcolsep}{1.8pt} % Default value: 6pt
   \renewcommand{\arraystretch}{1.0} % Default value: 1
	\begin{tabular}{l c c c c c c c c c c c c c} % four columns, alignment for each
     \hline
Id  &  MJD  &  $N_{\rm H1}$  & $C_{\rm f1}$   &  $N_{\rm H2}$ &  $C_{\rm f2}$ &$\Gamma$&$Norm^{\rm PL^\dagger}$&Fe $k_{\alpha}$ & EW&$Norm^{k_\alpha^\dagger}$   & $\rm log~L_{x}$ &$\rm log~\lambda_{ Edd}$ &$\chi^{2}/\rm dof$\\
    &        &  ($10^{22}$\cs)&     &   ($10^{22}$\cs) &     &      &   ($10^{-3}$)  & (keV)  &(eV) &  $(10^{-5})$ &$\rm log~(erg/s)$ &   &    \\
    &        &                &     &                  &     &      &     &    &      & 
    &  &   \\

\hline

XMM1 &51995&$ 1.97^{+0.26}_{-0.30}$& $ 0.92^{+0.01}_{-0.03}$&$ 6.21^{+2.16}_{-1.72}$&$ 0.54^{+0.09}_{-0.09}$& $ 1.53^{+0.10}_{-0.08}$ &$ 3.34^{+0.73}_{-0.57}$ &$  6.43^{+0.04}_{-0.05}$ &$  128^{+28}_{-26}$&$ 2.64^{+1.08}_{-0.83}$ &$ 43.06^{+0.03}_{-0.03}$ &$-1.83^{+0.02}_{-0.02}$&  764.06/741 \\
\\
XMM2 &53670& $ 1.92^{+0.25}_{-0.37}$& $0.94^{+0.01}_{-0.03}$&${ 5.26^{+2.20}_{-1.82}}$&$ 0.51^{+0.13}_{-0.11}$&$1.57^{+0.10}_{-0.10}$&$5.00^{+1.08}_{-0.86}$ &$ 6.40^{+0.04}_{-0.03}$ &$ 61^{+22.70}_{-23.20}$  & $ 1.75^{+0.64}_{-0.65}$ &$ 43.17^{+0.03}_{-0.04}$&$-1.70^{+0.02}_{-0.02}$   & 590.27/660  \\
\\
XRT1 &53754&$ 3.21^{+0.61}_{-1.28}$&$ 0.95^{+0.02}_{-0.07}$&$ 27.32^{+2.72}_{-2.14}$ & $ 0.62^{+0.03}_{-0.03}$&$ 1.73^{+0.10}_{-0.15}$&$ 6.94^{+2.3}_{-1.5}$&--&--&--&$ 42.81^{+0.06}_{-0.05}$&$-2.05^{+0.05}_{-0.05}$ & 119.97/116\\
\\
SU+NU1& 57133 & $ 4.97^{+3.38}_{-1.77}$ & $ 0.78^{+0.03}_{-0.03}$ & $ 30.68^{+4.29}_{-3.37}$ & $ 0.77^{+0.09}_{-0.03}$ & $1.72^{+0.06}_{-0.04}$ &$ 3.22^{+0.19}_{-0.17}$   &$ 6.39^{+0.06}_{-0.06}$ & $ 156^{+98}_{-75}$ &${1.78^{+0.37}_{-0.36}}$& $ 42.88^{+0.02}_{-0.02}$ &$ -2.08^{+0.01}_{-0.01}$& $ 364.51/389^{**}$\\
\\                  
XRT2+NU2 & 57335&$5.18^{+0.98}_{-0.89}$& $0.78^{+0.05}_{-0.04}$ & $ 14.24^{+1.22}_{-1.27}$ &$  0.87^{+0.02}_{-0.02}$ &$1.73^{+0.02}_{-0.02}$ &$ 2.98^{+0.02}_{-0.02} $   &$6.25^{+0.12}_{-0.09}$ &$103^{+30}_{-30}$  & $1.29^{+0.41}_{-0.49}$ & $43.03^{+0.05}_{-0.05}$ &$-1.87^{+0.02}_{-0.02}$ &$ 445.34/458^{**}$    \\
                     \\
XRT3a & 58647 & $  3.98^{+1.33}_{ -0.95}$  & $   0.92^{+0.02}_{ -0.02}$ &--& -- &$ 1.37^{+0.12}_{ -0.10}$ &$ 1.72^{+0.2}_{-0.2}$&--&--&--& $42.86^{+0.04}_{ -0.06}$           &$ -2.01^{+0.05}_{-0.05}$& 62.97/49 \\
\\
XRT3b & 58806 & $   2.72^{+0.74}_{ -0.80}$  & $  0.97^{+0.01}_{- 0.03}$&--&--&$  1.44^{+0.16}_{-0.15}$&$  3.06^{+1.20}_{-1.46}$&--&--&--& $ 42.89^{+0.08}_{ -0.08}$ &$ -1.98^{+0.08}_{-0.08}$           &  39/50\\
\\
XRT3c & 58832 & $  2.86^{+1.14}_{ -1.09}$  & $   0.91^{+0.05}_{ -0.12}$  &--&--&   $ 1.41^{ +0.16}_{ -0.17}$ &$ 1.10^{+0.13}_{-0.61}$&--&--&--& $ 42.79^{+0.03}_ {- 0.04}$&$ -2.12^{+0.04}_{-0.04}$ &  54.48/45\\
\\
XRT4a & 58892 & $ 4.07^{ + 1.47 }_{ - 1.46} $  & $  0.90^{ + 0.06}_{- 0.15} $ &--&--&  $  1.41^{+ 0.07}_{ - 0.06} $ &$0.84^{+0.11}_{-0.52}$  &--&--&--&$  42.54^{ + 0.06}_{- 0.06}$ &$ -2.34^{+0.06}_{-0.06}$&   59.63/45\\
\\
XRT4b & 59004 & $  5.34^{ + 1.75 }_{ - 1.60} $  & $  0.91^{ + 0.04}_{- 0.10} $ &--&--&   $  1.58^{+ 0.17}_{ - 0.15} $ &$0.91^{+0.88}_{-0.56}$&--&--&--&$  42.45^{ + 0.10}_{- 0.10} $& $ -2.40^{+0.10}_{-0.10}$&  32.95/36\\
\\
XRT4c & 59135 & $ 4.03^{ + 1.19 }_{ - 1.12} $  & $   0.93^{ + 0.03}_{- 0.07} $ &--&--&   $  1.39^{+ 0.13}_{ - 0.11} $ &$1.00^{+0.75}_{-0.50}$&--&--&--&$ 42.63^{ + 0.07}_{- 0.06} $&$-2.26^{+0.07}_{-0.07}$&    57.60/64\\
\\
XRT5a & 59271 & $  4.60^{ + 0.98 }_{ - 0.98} $ & $   0.94^{ + 0.02 }_{ - 0.03} $ &--&--&   $ 1.41^{+ 0.18}_{ - 0.23} $ &$ 1.29^{+0.47}_{-0.51}$&--&--&--&$ 42.76^{+ 0.03}_{- 0.03} $&$-2.14^{+0.03}_{-0.03}$&   82.27/94\\
\\
XRT5b & 59388 & $  3.68^{ + 0.60 }_{ - 0.60} $ & $  0.96^{ + 0.01 }_{ - 0.02} $ &--&--&   $  1.40^{+ 0.24}_{ - 0.23} $ &$ 1.96^{+1.00}_{-0.62}$&--&--&--&$  42.92^{+ 0.02}_{- 0.02} $&$ -1.97^{+0.02}_{-0.02}$&   148.68/143\\
\\
XRT5c & 59523 & $  3.45^{ + 0.68 }_{ - 0.64} $ & $   0.94^{ + 0.02 }_{ - 0.03} $ &--&--&   $  1.41^{+ 0.21}_{ - 0.23}  $ &$ 1.21^{+0.50}_{-0.43} $&--&--&--&$  42.77^{+ 0.02}_{- 0.02} $& $ -2.13^{+0.03}_{-0.03}$&  111.02/135\\
\\
XRT6a & 59641 & $  3.12^{ + 0.53 }_{ - 0.57} $ & $  0.97^{ + 0.01 }_{ - 0.02} $ &--&--&   $  1.43^{+ 0.20}_{ - 0.14}$ &$ 2.09^{+0.80}_{-0.30} $&--&--&--& $42.92^{+ 0.06}_{- 0.05} $&$ -1.97^{+0.03}_{-0.03}$&  125.27/132\\
\\
XRT6b & 59752 &  $  2.48^{ + 0.46 }_{ - 0.51}$  & $  0.96^{ + 0.01 }_{ - 0.01} $ &--&--&    $  1.41^{+ 0.22}_{ - 0.23} $ & $ 1.52^{+0.67}_{-0.46} $ &--&--&--&$  42.81^{+ 0.03}_{- 0.03} $& $ -2.07^{+0.03}_{-0.03}$&  119.29/127\\
\\
XRT6c & 59886 & $ 3.24^{ + 0.62 }_{ - 0.62} $ & $   0.95^{ + 0.03 }_{ - 0.03} $ &--&--&   $  1.49^{+ 0.20}_{ - 0.26} $&$ 1.66^{+0.57}_{-0.52}$ &--&--&--&$ 42.73^{+ 0.04}_{- 0.03}$ & $ -2.16^{+0.03}_{-0.03} $ &   99.05/115 \\
\\                 
\hline
	\end{tabular}
 \leftline{$^\dagger$ In the unit of \normflux}
 \leftline{** The fit statistics for each instrument in the broadband fitting SU+NU1 is: 
106.26/69 for XIS, 133.69/162 for FPMA, and 124.56/158 for FPMB.}
\leftline{Similarly, in XRT2+NU2 broadband fitting, fit statistics is: 104.48/124 for XRT, 191.71/166 for FPMA, and 149.15/168 for FPMB.}
 \end{table*}
\endgroup

\begin{figure*} 
	\centering
	\includegraphics[scale=1.3]{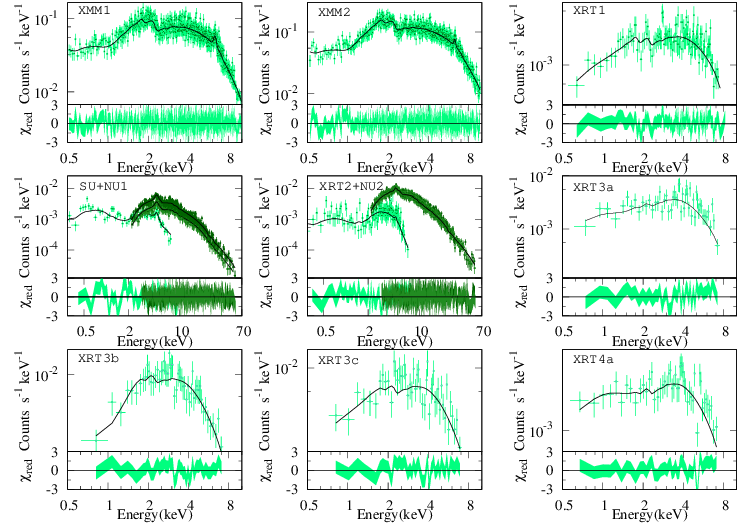}
    \includegraphics[scale=1.3]{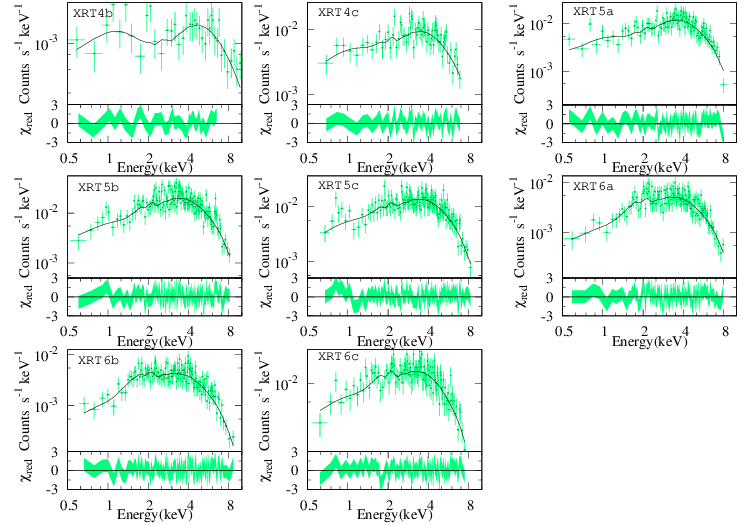}
	\caption{{\tt Powerlaw } model fitted spectra of Mrk~6 from the XMM–Newton, Suzaku, NuSTAR, and Swift observations along with the residuals obtained from the
spectral fitting. }
	\label{fig:all_plotsA} 
\end{figure*}

In the subsequent analysis, we use the data from two broad-band observations, namely, SU+NU1 (MJD--57133) and XRT2+NU2 (MJD--57335), in the energy range of 0.5--60 keV. The parameters obtained after applying the best-fitted composite model to the spectra are  $\Gamma$ = 1.72$\pm$0.05 and 
1.73$\pm$0.02, iron $\Kalpha$ line at $6.39\pm0.06$ and $6.25\pm0.10$ keV with EW of $ 156^{+98}_{-75}$ and $103^{+30}_{-30}$, respectively. The broad-band spectra fitted with this composite model for the SU+NU1 and XRT2+NU2 observations are shown in Figure~\ref{fig:all_plotsA}. Next, we replaced the second 
{\tt pcfabs} component with {\tt zxipcf} to check the ionization of the absorber. The best-fitting values obtained in the presence of an ionized absorber are reported in Table~\ref{tab:zxipcf}.

Applying the same baseline model to the remaining  12 \Swift~ observations, we find that the double \texttt{pcfabs} does not significantly improve the fit. This could be due to the low exposure time of the observations, relatively poor energy resolution of XRT, or the absence of the second absorber. Consequently, we opt to substitute the double \texttt{pcfabs} with a single \texttt{pcfabs} from XRT3a (MJD-58647) to XRT6c (MJD-59886). In these cases, our chosen baseline model becomes ${\tt TBabs\times pcfabs \times powerlaw}$.  Across these observations, the average photon index ($\Gamma$) is estimated as 1.43. For the single absorber, the absorption column density and corresponding covering factor vary in the ranges of ($ 2.48^{+0.46}_{-0.51}$ -- $5.34^{+1.75}_{-1.60})\times10^{22}~\text{cm}^{-2}$, and ($0.90^{+0.06}_{-0.15}$ -- $0.97^{+0.01}_{-0.03})$, respectively. The X-ray continuum luminosity of the source in the 3.0--10.0 keV energy band, estimated from the power-law fitting to the data from these observations, varies in the range of $ 42.45^{+0.10}_{-0.10}$ to $ 42.92^{+0.02}_{-0.02}$, whereas the Eddington ratio $\rm log(\lambda_{Edd})$ varied in the range of  -2.40 to -1.97. The details of the best-fitted results are given in Table~\ref{tab:powerlaw}.

\begin{figure*} 
	\centering
	\includegraphics[trim={0 1cm 0cm 0},clip,scale=1.3]{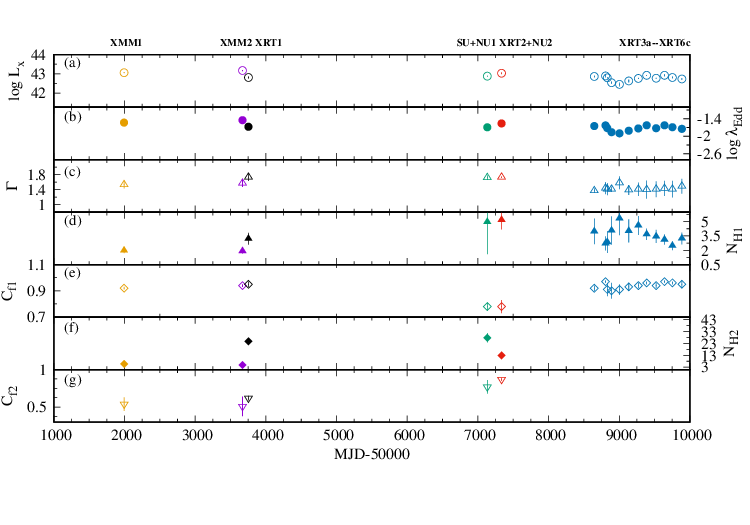}
\caption{Temporal variations of X-ray luminosity ($L_x$), Eddington ratio ($\lambda_{\rm Edd}$), photon index ($\Gamma$), hydrogen column density and corresponding covering factor for two distinct absorbers ($N_{\rm H1}$ \& $C_{\rm f1}$, and $N_{\rm H2}$ \& $C_{\rm f2}$) are shown for all the observations used in the present work. The values of $L_x$ and  $\lambda_{\rm Edd}$  are plotted in logarithmic scale. The values of the parameters are also given in Table~\ref{tab:powerlaw}. }
	\label{fig:Parameter} 
\end{figure*}

The luminosity variation in X-ray band ($L_x$), Eddington ratio ($\lambda_{\rm Edd}$), photon index ($\Gamma$), hydrogen column densities for the two distinct absorbers ($N_{\rm H1}$ \& $N_{\rm H2}$), and corresponding covering factors ($C_{\rm f1}$ \& $C_{\rm f2}$) are graphically presented in Figure~\ref{fig:Parameter}, for all the observations used in the present work. The X-ray luminosity of the source remains comparable throughout our observational period and is presented in panel (a) of Figure~\ref{fig:Parameter}. The average value of $\rm log(L_x)$ is $ 42.82^{+0.07}_{-0.07}$, with the highest and lowest values being $ 43.17^{+0.01}_{-0.01}$ and $ 42.45^{-0.10}_{+0.10}$, respectively. The Eddington ratio of Mrk~6 is shown in panel (b) of Figure~\ref{fig:Parameter}. 
 
However, we observe that the photon index ($\Gamma$) varies between $ 1.37^{+0.12}_{-0.10}~\text{and}~ 1.73^{+0.02}_{-0.02}$. The variation of $\Gamma$ is presented in panel (c) of Figure~\ref{fig:Parameter}. We observe that Mrk~6 has a complex column density structure around it. In the initial phase of our observations, two-column densities of different nature are used to describe the absorption. However, towards the later part of our observation (from XRT3b), the complexity in the absorption disappears. The variation of hydrogen column density with the corresponding covering factor for both the absorbers are shown in panels (d,  e) and (f,  g) of Figure~\ref{fig:Parameter}, respectively.  Figure~\ref{fig:all_plotsA} shows the best-fitting spectrum obtained with the {\tt powerlaw} model.

\begingroup
\begin{table*}
\centering
\caption{Best-fitting parameters obtained from the spectral fitting of the data with Model: ${\tt TBabs\times pcfabs\times zxipcf\times (powerlaw+zGauss)}$. The ionizing luminosity $\rm L_{ion}$ is calculated in the energy range of 13.6 eV--13.6 keV (1--1000 Ryd).}
\label{tab:zxipcf}
\setlength{\tabcolsep}{4.0pt} % Default value: 6pt
\renewcommand{\arraystretch}{1.0} % Default value: 1
\begin{tabular}{l c c c c c c c c c c c c c c} % four columns, alignment for each
\hline
Id  &  MJD  &  $N_{\rm H1}$  & $C_{\rm f1}$   &  $N_{\rm H2}$ & $\rm log\xi$& $C_{\rm f2}$   &$\Gamma$ &$Norm^{\rm PL}$ &  $\chi^{2}/\rm dof$&$ log L_{ion}$&$ r_{max}$\\
    &       &   $(10^{22}$\cs) &      & $(10^{22}$\cs)  &   &    &    &$(10^{-3})$ & & $\rm log~(erg/s)$ & pc \\
    &       &                  &      &                 &  &  &    &    &\normflux &     \\
\hline

XMM1      &  51995  &  $ 1.31^{+0.57}_{-0.38}$  &  $ 0.91^{+0.05}_{-0.06}$   &  $ 6.47^{+2.65}_{-1.10}$  & $ 1.08^{+0.23}_{-0.26}$  & $ 0.70^{+0.14}_{-0.17}$  &    $ 1.58^{+0.08}_{-0.08}$ &  $3.48^{+0.27}_{-0.28}$ &  763.02/740&$43.51^{+0.03}_{-0.02}$  &$ 13.50$\\
\\
XMM2      &  53670  &  $ 1.76^{+0.21}_{-0.50}$  &  $0.93^{+0.05}_{-0.05}$   &  $ 5.83^{+1.07}_{-1.01}$  &  $ 1.20^{+0.20}_{-0.19}$& $ 0.67^{+0.17}_{-0.18}$  &    $ 1.56^{+0.10}_{-0.09}$ &  $ 5.55^{+1.30}_{-0.40}$ & 595.08/658&$43.65^{+0.06}_{-0.04}$ & $ 15.69$\\
\\
XRT1      &  53754  &  $1.95^{+1.01}_{-1.02}$  &  $ 0.85^{+0.05}_{-0.04}$   &  $ 6.69^{+4.52}_{-8.39}$ &  $1.05^{+0.15}_{-0.19}$ & $0.80^{+0.04}_{-0.03}$  &    $ 1.71^{+0.10}_{-0.15}$ &  $ 4.95^{+0.27}_{-0.28}$ &  118.81/115&$43.58^{+0.02}_{-0.04}$   &$16.44$\\
\\
SU+NU1    &  57133  &  $ 2.76^{+0.52}_{-0.92}$  &  $ 0.71^{+0.04}_{-0.04}$   &  $ 12.91^{+3.40}_{-11.10}$ &  $ 1.03^{+0.16}_{-0.30}$ & $ 0.81^{+0.01}_{-0.02}$  &    $1.72^{+0.10}_{-0.09}$ &  $ 3.00^{+0.10}_{-0.08}$  &$ 357.49/388^{**}$&$43.25^{+0.03}_{-0.06}$   & $ 4.17$\\
\\
XRT2+NU2  &  57335  &  $ 7.70^{+1.03}_{-1.90}$  &  $ 0.88^{+0.02}_{-0.05}$   &  $ 12.65^{+7.75}_{-8.20}$ &  $ 1.10^{+0.44}_{-0.29}$ & $ 0.92^{+0.02}_{-0.02}$  &    $1.73^{+0.07}_{-0.10}$ &  $4.86^{+0.87}_{-0.71}$ & $ 442.03/457^{**}$&$43.37^{+0.03}_{-0.05}$   & $ 4.78$\\
\\

\hline
		
\end{tabular}
\leftline{** The fit statistics for each instrument in the broadband fitting SU+NU1 is: 
96.67/68 for XIS, 135.11/162 for FPMA, and 125.71/158 for FPMB.}
\leftline{Similarly, in XRT2+NU2 broadband fitting, fit statistics is: 104.67/123 for XRT, 188.73/166 for FPMA, and 148.63/168 for FPMB.}
\end{table*}
\endgroup

\subsubsection{\textbf{\tt nthcomp} Model}
As the power-law model effectively fits the primary continuum in the 3.0 -- 10.0 keV range, we attempt to estimate the temperature of the Compton cloud. Hence, we replace the {\tt powerlaw} with {\tt nthcomp} model \citep{Zdziarski1996,Zycki1999}. The thermally Comptonized continuum model is parameterized by the  hot  electron temperature $ kT_e$. This model depends on the energy of the seed photons $(kT_{\rm bb})$, which we consider at 10 eV for all the observations. It is to be noted that we vary this parameter from 1 eV to 50 eV and do not notice any variation in the residuals of the fitted spectra. Alongside this, we consider disc-blackbody type seed photons for this work. To do so, we opt {\tt int-type=1} for all the spectral fitting. The {\tt nthcomp} model provides us the photon index ($\Gamma$) and the hot electron temperature of the Compton cloud ($ kT_e$). Furthermore, we calculate the optical depth ($\tau$) for each observation using the formula :

\begin{equation}
\tau=\sqrt{\frac{9}{4}+\frac{3}{\theta_{e}(\Gamma+2)(\Gamma-1)}}-\frac{3}{2}
\label{eq:tau}
\end{equation}
by using the relation as presented in \cite{Zdziarski1996}, where $\theta_{e}=\frac{kT_{e}}{m_{e}c^{2}}$ is the electron energy with respect to the rest mass energy. 

Simultaneous broadband observations of Mrk~6 were carried out with {\it Suzaku} and \nustar, and {\Swift} and \nustar, at two epochs, separated by 202 days. While fitting the broadband spectra for both the epochs with the {\tt nthcomp} model, we find that the photon index ($\Gamma$) remains almost constant ($\Gamma\sim 1.70-1.72$).
This suggests that the nature of the Compton cloud remained stable over this period. The electron cloud temperature ( $kT_{e}$) is found to exceed 65 keV for SU+NU1 observation and is determined as $ 69^{+15}_{-9}$ keV for XRT2+NU2 observation. In both cases, the values of the optical depth ($\tau$) remain almost constant ($\tau \sim 1.83-1.71$). The results obtained from our spectral fitting with this model are listed in Table~\ref{tab:nthcomp}. We also include a Gaussian component to account for the Fe-line near $\sim6.4$ keV, and the results closely resemble those obtained from the power-law fitting.

\begin{table*}

	\centering
	\caption{{\tt nthcomp} model fitting results for the spectrum above 3.0 keV. The optical depth $\tau$ is calculated from Equation~\ref{eq:tau}}
	\label{tab:nthcomp}
	\begin{tabular}{l c c c c c c } % four columns, alignment for each
     \hline
Id  &  MJD  &  $\Gamma$ & $kT_{e}$    &  $Norm^{\rm nth}$ & $\chi^{2}/\rm dof^{**}$ & $\tau^*$\\
    &       &                &  (keV)& $(10^{-3})$ &     &    \\
     &       &                &  & \normflux &     &    \\
     \hline
SU+NU1    &  57133 & $ 1.70^{+0.09}_{-0.10}$   &  $ >65$ &  $ 3.52^{+0.31}_{-0.30}$    &   305.43/363  & $ <1.83$  \\
\\
XRT2+NU2  &  57335 &  $ 1.72^{+0.05}_{-0.07}$   &  $ 69.51^{+14.82}_{-9.46}$ &  $2.98^{+0.36}_{-0.23}$   &   416.06/419 & $ 1.71^{+0.23}_{-0.58}$\\ 
\\
\hline
	\end{tabular}
\leftline{** The fit statistics for each instrument in the broadband fitting SU+NU1 is: 
48.04/45 for XIS, 132.88/161 for FPMA, and 124.51/157 for FPMB.}
\leftline{Similarly, in XRT2+NU2 broadband fitting, fit statistics is: 69.13/85 for XRT, 191.81/166 for FPMA, and 155.12/168 for FPMB.}

\leftline{$^*$ The optical depth is not estimated from the spectral fitting. It is calculated using Equation~\ref{eq:tau}.}
 
\end{table*}

\subsubsection {\textbf{\tt pexrav} Model}
\label{sec:pexrav}
Although the power-law model provides a satisfactory fit of the primary continuum in the 3.0--10.0 keV energy range, these models fail to characterize the high energy (above 10.0 keV) counterpart of the observed spectrum of Mrk 6. So, we use another power law, which is found to be flattened compared to the continuum.  Moreover, the Fe line EW of more than 200 eV is a signature of the presence of a reflection component from an obscured medium \citep{Krolik1994}. The flatness of the second power-law component, along with the high equivalent width of the observed Fe K$_\alpha$ line, suggests the presence of a reflection component in the broadband spectrum of Mrk~6. To address the reflection component, we substitute the second power law with the slab reflection model {\tt pexrav} \citep{Zdziarski1996}. Consequently, the composite model utilized to fit the broadband spectra of both the epochs of observations is as follows:

\begin{center}
\begin{multline*}
    {\tt TBabs\times pcfabs\times zxipcf\times const(cutoffpl+zGauss+pexrav)}.
\end{multline*} 
\end{center}

The two column densities, namely {\tt pcfabs and zxipcf}, are used to incorporate absorption along the line of sight (for more details, see Section~\ref{sec:pl}). Regarding the model {\tt pexrav}, we tie the photon index with the cut-off power-law index $(\Gamma)$ while keeping this $\Gamma$ as a free parameter. As the cut-off power-law model is used as the primary continuum, we constrain $rel_{refl}<0$, indicating that {\tt pexrav} solely contributes to the reflection component only. We fix the abundance to the solar value  while the inclination angle ($i$) and reflection fraction ($R_{\rm f}$) are kept free during spectral fitting.

 For the SU+NU1 (MJD--57133) observation, we find the photon index and cut-off energy at $\Gamma=1.72^{+0.10}_{-0.09}$ and $E_{\rm c}=111^{+39}_{-28}$ keV, respectively. The {\tt pexrav} component of the composite model gives the value of the reflection coefficient as $ R_{\rm f}=1.85^{+0.16}_{-0.15}$. Subsequently, in the case of the XRT2+NU2 (MJD--57335) 
observation, we find $\Gamma=1.75^{+0.07}_{-0.10}$, cut-off energy $E_{\rm c}=121^{+38}_{-35}$ keV and reflection fraction $R_{\rm f}=1.02^{+0.15}_{-0.18}$. The comparable parameters obtained during the two epochs of observations may be due to the 202-day gap between the epochs, during which the source properties did not show any noticeable change in its properties. It is to be noted that the parameter $i$ is insensible during these spectral fittings. We obtained that $i = 63\pm 45$ degrees for both cases.

\begingroup
\begin{table*}
	\centering
	\caption{The parameters obtained from the spectral fitting of broadband (0.5 to 60.0 keV) data of Mrk~6 with the model: ${\tt TBabs\times pcfabs\times zxipcf\times const(cutoffpl+zGauss+pexrav)}$. } 
	\label{tab:pexrav}
    \setlength{\tabcolsep}{2.8pt} % Default value: 6pt
   \renewcommand{\arraystretch}{1.0} % Default value: 1
	\begin{tabular}{l c c c c c c c c c c c   } % four columns, alignment for each
     \hline
Id  &  MJD  &  $N_{\rm H1}$  & $C_{\rm f1}$   &  $N_{\rm H2}$ & $C_{\rm f2}$&$\rm log~\xi$&    $\Gamma$ & $E_{\rm C}$&$R_{\rm f}$&$Norm$ &  $\chi^{2}/\rm dof^{**}$\\
    &       &   $(10^{22}$\cs) &      & $(10^{22}$\cs)  &    &    & & (keV) &     & $(10^{-3})$&  \\
    &       &                  &      &                 &    &    &        &      & &\normflux    &    \\
\hline
SU+NU1    &  57133  &  $ 3.20^{+0.52}_{-0.57}$  &  $0.49^{+0.18}_{-0.19}$   &  $ 17.56^{+1.15}_{-2.45}$ &   $ 0.91^{+0.18}_{-0.14}$  &  $ 0.98^{+0.11}_{-0.51}$   &  $1.72^{+0.10}_{-0.09}$ & $111^{+39}_{-28}$  & $1.85^{+0.16}_{-0.15}$  &$ 2.41^{+0.70}_{-0.47}$& $ 751.16/814$\\
\\
XRT2+NU2  &  57335  &  $3.36^{+0.55}_{-0.64}$  &  $0.57^{+0.04}_{-0.04}$   &  $12.60^{+5.14}_{-5.41}$ &   $0.96^{+0.22}_{-0.29}$   &  $0.82^{+0.12}_{-0.15}$   &    $1.75^{+0.07}_{-0.10}$ & $121^{+38}_{-35}$ &$1.02^{+0.15}_{-0.18}$   &  $4.86^{+0.87}_{-0.71}$ & $ {889.50/881}$\\
 \\ 
  \hline
	\end{tabular}
 \leftline{** The fit statistics for each instrument in the broadband fitting SU+NU1 is: 
102.97/71 for XIS, 338.32/381 for FPMA, and 309.87/362 for FPMB.}
\leftline{Similarly, in XRT2+NU2 broadband fitting, fit statistics is: 118.25/130 for XRT, 443.04/381 for FPMA, and 328.21/370 for FPMB.}
\end{table*}
\endgroup

\begingroup
\begin{table*}
	\centering
	\caption{The parameters obtained from the spectral fitting of broadband (0.5 to 60.0 keV) data of Mrk~6 with the model:  ${\tt const1\times Tbabs\times (atable\{borus02\}+zphabs\times cabs\times cutoffpl1+const2\times cutoffpl2)}$.}
 \label{tab:borus}
 \setlength{\tabcolsep}{4.0pt} % Default value: 6pt
   \renewcommand{\arraystretch}{1.0} % Default value: 1
	\begin{tabular}{l c c c c c c c c c c c  } % four columns, alignment for each
    \hline 
   Id  &  MJD &$\Gamma^{\rm BR}$&$E_{\rm cut}$&$N^{\rm Tor}_{\rm H}$ & $C^{\rm Tor}_{\rm f}$ & $\theta^{\rm tor}$& $i$ & $A_{\rm Fe}$ & $N^{\rm los}_{\rm H}$&$Norm^{\rm BR}$   & $\chi^{2}/\rm dof^{**}$\\
       &      &            & (keV) & ($10^{25}$\cs)   &    &  Degree& Degree& ($A_{\odot}$) & ($10^{22}$\cs)&$(10^{-3})$  &     \\
   
       &      &             &  &    &    &   &    &      &  &  \normflux     &     \\
       \hline
 SU+NU1   &  57133   & $1.72^{+0.09}_{-0.12}$ & $ 105^{+34}_{-17}$ & $1.41^{+1.35}_{-1.32}$ & $0.80^{+0.26}_{-0.28}$ &$36.74^{+3.78}_{-3.78}$ & $18.95^{+3.70}_{-3.91}$ & $0.24^{+0.08}_{-0.09}$ & $13.16^{+1.50}_{-1.50}$ &$ 2.48^{+0.36}_{-0.30}$  &   $ 786.72/811$      \\
\\                     

 XRT2+NU2 &  57335   &$ 1.77^{+0.10}_{-0.13}$ & $120^{+22}_{-18}$ &$ 1.02^{+1.12}_{-1.10}$ &$0.80^{+0.22}_{-0.29}$ &$37.07^{+3.95}_{-4.98}$ &$19.05^{+4.99}_{-4.19}$  &$0.21^{+0.08}_{-0.08}$&  $15.54^{+1.75}_{-1.75}$ & $ 5.30^{+0.58}_{-1.71}$& $ 883.13/876$ \\
\\
  \hline
	\end{tabular}
 \leftline{** The fit statistics for each instrument in the broadband fitting SU+NU1 is: 
110.80/70 for XIS, 365.57/380 for FPMA, and 310.35/361 for FPMB.}
\leftline{Similarly, in XRT2+NU2 broadband fitting, fit statistics is: 95.16/124 for XRT, 458.35/383 for FPMA, and 329.62/369 for FPMB.}
  
\end{table*}
\endgroup

\subsubsection {\textbf{\tt Borus Model}}
\label{sec:borus}

Although the disc-reflection model provides an acceptable fit, it is worth noting that the {\tt pexrav} model is inadequate to explain the properties of the reflective medium in detail. Indeed, this reflection model offers valuable insights into the temperature of the Compton cloud and the reflection coefficient. However, it is important to highlight that {\tt pexrav} assumes an infinite line-of-sight column density and overlooks the finite characteristics of the reflecting medium. Therefore, in the subsequent analysis, we opt for a more comprehensive toroidal reflection model known as {\tt borus} \footnote{\url{https://sites.astro.caltech.edu/~mislavb/download/}} \citep{Balokovic2018}. This model assumes a spherical reprocessing medium with a conical cut-out at the poles. This provides a toroidal structure of the reflecting materials, exhibiting a variable covering factor. For this work, we use {\tt borus02}\footnote{\tt $borus02\_v170323c.fits$} model, which offers supplementary information on parameters such as the cut-off energy $(E_{\rm cut})$ and abundance of Fe $(A_{\rm Fe})$ in addition to other model parameters. The composite model employed to fit the broadband spectra using {\tt borus02} is described as follows:

\begin{center}
\begin{multline*}
    {\tt const1\times Tbabs\times (atable\{borus02\}} \\ {\tt+zphabs\times cabs\times cutoffpl1+const2\times cutoffpl2)}.
\end{multline*}
\end{center}

$ {\tt zphabs\times cabs\times cutoffpl1}$ represents the absorbed direct primary emission while ${\tt const2\times cutoffpl2}$ represents the scattered primary emission. In the above model, {\tt const1} and {\tt const2} are instrument cross-normalization and the relative normalization of a leaked or scattered unabsorbed reﬂection of the intrinsic continuum, respectively. 
We allow the torus column density ($N_{\rm H}^{\rm tor}$) with the toroidal angle ($\theta^{\rm tor}$) and the inclination angle ($i$) to vary freely during the broadband spectral fitting, as these parameters are unknown to us. For other model parameters, we follow the methodology outlined in \cite{Balokovic2018}, and the corresponding outcomes are presented in Table~\ref{tab:borus}.

The column density $N_{\rm H}$ for {\tt zphabs} is kept unconstrained during the broadband spectral fitting. Furthermore, we tie the normalization values, cut-off energies, and the photon indices for both the cut-off power-law models ({\tt cutoffpl1} and {\tt cutoffpl2}) to the {\tt borus02} model.  The column densities of {\tt cabs} and {\tt zphabs} models are tied together and represent the line-of-sight absorption at the redshift of the source.

 From the broadband spectral fitting of SU+NU1 and XRT2+NU2 observations with {\tt borus02} model,  the estimated model parameters align consistently with the values found in spectral fitting using other models. For instance, the average photon index ($\Gamma$) and the cut-off energy $E_{\rm cut}$) obtained from {\tt borus02} fitting, approximately 1.7 and 115 keV, respectively, are very similar to the results obtained from fitting with the {\tt nthcomp} model. Additionally, this model ({\tt borus02}) provides insights into the properties of the torus derived from the X-ray spectral fitting. We find the average hydrogen column density of the torus $N_{\rm H}^{\rm tor}=1.2\times10^{25}$ ~cm$^{-2}$ with half-opening angle $\theta_{\rm tor}=37\pm4$ degree and a constant covering factor $C^{\rm tor}_{\rm f}=0.8\pm0.3$ for these observations. Furthermore, the {\tt borus02} model fitting for SU+NU1 and XRT2+NU2 observations yield an inclination angle of $ i\sim 19\pm4$ degrees and iron abundances of $A{\rm Fe}\sim 0.2\pm0.08$. The detailed results obtained from this fitting are reported in Table~\ref{tab:borus}.

\section{Discussion}
\label{sec:discussion}
In this study, we investigated long-term X-ray temporal and spectral properties of Mrk~6 using data from various X-ray missions, such as \xmm, \nustar, \Swift, and {\it Suzaku}.  We used various phenomenological ({\tt powerlaw}, {\tt cutoffpl} and  {\tt pexrav}) and physical ({\tt nthcomp} and {\tt borus}) models to study the X-ray emitting region throughout our observational period. In this section, we discuss the key findings of the above analysis.

\subsection{Evolution of primary continuum}
 The primary continuum in the X-ray energy band of AGNs is believed to arise through the process of inverse Comptonization in a hot electron cloud $(\sim 10^9 K)$, called Compton cloud or corona \citep{ST1980}. The UV/optical seed photons from the accretion disk \citep{SS1973} are up-scattered in this Compton cloud and produce X-ray photons. It is believed that this hot electron cloud is located in the nearby region of the central black hole. However, the geometry of the Compton cloud is still a topic of ongoing research. 

Mrk~6 is a nearby,  relatively unexplored AGN that has exhibited ``changing-look'' behaviour in optical observations \citep{Osterbrock1976, 1983ApJ...265...92M, 2003ASPC..290...89D, 2014MNRAS.440..519A,2019sf2a.conf..509M, 2022ApJ...927..227L}. However, its characteristics in the X-ray band have not been thoroughly investigated. As we have a very limited understanding of its spectral behavior in X-rays, our primary motivation is to explore various properties of the Compton cloud.  As the primary continuum is well-fitted by the power-law model, we can explain the variation in the primary continuum by studying the variability in the photon index ($\Gamma$). Throughout our observations between 2001 and 2022, we noticed a minor variation in $\Gamma$ of the primary continuum. In the 2001 and 2005 observations, we observed a similar type of continuum with $\Gamma~\sim~1.5\pm0.1$. However, the Eddington ratio increased in the 2005 observation. 

 During the 2006 observation, we observed a steepened spectrum with $\Gamma=1.73\pm0.1$, indicating that the source transitioned to a softer state compared to the previous observations. The estimated source luminosity of $10^{42.81}$ erg/s is approximately 2.3 times less than that during the 2005 observation. This change in the spectral state suggests a shift in the emission properties of the source, potentially indicating alterations in its accretion dynamics and radiation mechanisms.

 In the 2015 observations, we identified a comparable X-ray continuum to that observed in 2006, reflected in a similar spectral index of $\Gamma=1.7\pm0.1$ and an average X-ray luminosity of $10^{42.9}$ erg/s. These observations, complemented by a high-energy counterpart from \nustar, allowed us to estimate the high energy cut-off, yielding an average value of $E_c\sim115\pm40$ keV. Corresponding electron temperature $(kT_e)$ obtained from spectral fitting is $kT_e>65$ keV. We also estimated the optical depth $(\tau)$ for these observations to be $\tau<1.83$. 

 Following the broadband observations in 2015, \Swift~ continued to observe the source from 2019 to 2022, segmented into twelve binned observations. Our spectral analysis across this period revealed minimal variation in the X-ray continuum. The average photon index $(\Gamma)$ of the power-law continuum stands at $1.43\pm0.05$. The highest photon index $(\Gamma=1.58\pm0.17)$, indicating the softest spectrum, was observed in 2020 (XRT~4b). Conversely, the hardest spectrum, denoted by the lowest value of $\Gamma=1.37\pm0.12$, was observed in 2019 (XRT~3a). The luminosity and the  Eddington ratio vary accordingly.

Along with the primary X-ray continuum, a narrow Fe-line was evident in observations from \xmm, \suz, and \nustar. However, due to the spectral resolution and exposure time limitations of \Swift, we failed to detect any line in the \Swift~ spectra. To check the presence of Fe-line robustly, we combined all \Swift~ spectra into a single spectrum. The Fe-line remains undetected even after combining. From the spectral fitting of data from the \xmm~\& ~\nustar~observations, we identified the Fe-line at approximately $\sim6.4$ keV, exhibiting variations in equivalent width (EW). The average EW measured was $112\pm45$ eV, with the highest recorded value of $156^{+98}_{-75}$ eV observed in the 2015 dataset, while the lowest was noted as $61\pm23$ eV in the 2005 observation.

Based on the overall results obtained from X-ray spectroscopy, we found that the nature of the Compton cloud changed with time. The spectral state transitioned from a relatively harder state $(\Gamma\sim1.5)$ to a comparatively softer state  $(\Gamma\sim1.7)$ and again became harder $(\Gamma\sim1.4)$ towards the end of the observation period. We observed that the Eddington ratio $\rm log(\lambda_{Edd})$ varied from $-1.70$ to $-2.40$ with an average of $-2.04$. This indicates that the source remained in the sub-Eddington regime during these 22 years of the observations.

\subsection{Properties of the absorbing medium}

Earlier observations across various energy bands (from radio to X-ray) pointed out that Mrk~6 has a complex gas structure around the central X-ray emitting region. \cite{Feldmeier1999} provided the first insight into the complexities of the gas structure. They also proposed that the X-ray photons might be influenced by gas containing trace amounts of dust, or this gas could be situated within the Broad Line Region (BLR). Later, using BeppoSAX (1999), ASCA (1997), and \xmm~(2001) observations, \cite{Immler2003} reported that Mrk~6 favors a double partial-covering model consisting of partially ionized and neutral gas along the line of sight. They suggested that the observed variability could be attributed to the dynamic gas movements within the torus. However, \cite{Schurch2006} concluded from the 2003 \xmm~ observation that the absorbing gas along the line of sight exhibited characteristics of the outflow. This idea found support in previous radio studies \citep{Capetti1995} and spectro-polarimetric observations \citep{Khachikian2011}.

Initially, it became apparent that this source exhibited a double partially covering hydrogen column structure along the line of sight. However, as we progressed, a single partially covered absorption model was sufficient to describe absorption in the X-ray spectrum for observations from 2019 to 2022. This intriguing transition indicates that the gas structure surrounding the central engine was complex until 2015. After that, this complexity appeared to diminish after 2015.  We find the  value of $N_{\rm H1}$ at $1.9\times10^{22}~\text{cm}^{-2}$ for the 2001 and 2005 \xmm~ observations. However, after that, $N_{\rm H1}$ increased and reached $ 5.07\times10^{22}~\text{cm}^{-2}$ during the 2015 broadband observation. Later, the neutral hydrogen column density $(N_{\rm H1})$ became nearly constant at $ 3.50\times10^{22}~\text{cm}^{-2}$ for the rest of the observations. From the above discussion, we infer that the neutral hydrogen column density in Mrk~6 remains relatively consistent in our observational period (2001--2022) at around $ (2-5)\times10^{22}~\text{cm}^{-2}$ with a small variation within uncertainty. With this, the covering factor of this component remains consistently high, approximately $>90\%$ for most of the observations. These findings imply that this neutral hydrogen cloud likely resides at a considerable distance from the central engine, potentially near or above the torus and along the line of sight.

On the other hand, we observed another type of hydrogen cloud, which was ionized, during the X-ray spectral fitting below 3.0 keV. Initially, we modeled this using a simple model known as {\tt pcfabs} (see Section~\ref{sec:pl}). However, we later replaced it with the more sophisticated {\tt zxipcf} model. We found that the column density $(N_{\rm H2})$ and the covering factor $(C_{\rm f2})$ of this component vary with time. For 2001 \xmm~ observation, we encountered $ (N_{\rm H2}\sim6.2\times10^{22}~\text{cm}^{-2})$, which later increased and reached at $ (N_{\rm H2}\sim31\times10^{22}~\text{cm}^{-2})$ during the simultaneous observations in 2015. Beyond 2015, this component of hydrogen column density seemed to dissipate. Furthermore, we observed that the covering factor increased over time. In 2001, this $N_{\rm H2}$ covered nearly $\sim 54\%$ of the X-ray emitting region. However, the value of this parameter increased to $\sim 78\%$ during 2015 observations. We also estimated the ionization parameter $(\log\xi)$ for these observations and found that this parameter remained relatively stable with  $\log\xi \sim  1-1.2$. Given the ionized nature of this medium, exhibiting temporal variations, it may be inferred that this ionized hydrogen cloud resides comparatively close to the X-ray-emitting region. 

 To calculate the maximum possible distance of this ionized cloud, we used the formula $r \leq r_{\rm max} = \frac{L_{\rm ion}}{N_{\rm H}\xi}$ \citep{Blustin2005, Crenshaw2012}. Utilizing the ionizing luminosity ($L_{\rm ion}$) calculated in the energy range of 13.6 ev to 13.6 keV at redshift (z) of 0.0186, and $N_{\rm H}$, we calculated the maximum radial distance, $r_{\rm max}$ (see Table~\ref{tab:zxipcf}). Initially, the ionized absorber extended up to 13.50 pc during the 2001 \xmm~ observation, expanded further to 15.69 pc in 2005, and subsequently decreased. The SU+NU1 and XRT2+NU2 broadband observations indicated a cloud spanning 4.17 and 4.78 pc, respectively. As the outer radius moved inward, we observed a higher hydrogen column density for these observations: $12.91^{+3.40}_{-11.10}\times10^{22}$cm$^{-2}$ and $12.65^{+7.75}_{-8.20}\times10^{22}$ cm$^{-2}$, respectively, while the ionization remained comparable to the previous observations. Our findings suggest that the ionized cloud, initially at 15.69 pc, moved inward to 4.76 pc. These locations potentially coincide with AGN components like the Narrow Line Region (NLR) or torus \citep{Kaastra2012, Reeves2013, Laha2016}. The dissipation or shift in the position of this region may related to the outflow as suggested in previous studies \citep{Capetti1995, Kharb2006, 2014MNRAS.440..519A}.

 The analysis of broadband X-ray data (0.5 to 60.0 keV range) using the 'borus02' model provided insights into the characteristics of the obscuring materials around the central engine. From the spectral fitting, we found that the average hydrogen column density ($N_{\rm H}^{\rm Tor}$) is nearly constant at $\sim1.21\times10^{25}~\text{cm}^{-2}$ with approximately $\sim80\%$ covering and toroidal opening angel from pole $\theta^{\rm tor}\sim37$ degrees (see Section~\ref{sec:borus}). The Fe abundance in the torus is also found to be nearly constant at $\sim0.22~A_{\odot}$. We found the value of the inclination angle $i$ nearly constant at $19.0^\circ$. This stability observed in the nature of the source could be due to the relatively short time gap between these two observations, which occurred within 202 days.

From the above study on Mrk~6 in the X-ray band, we unveiled a complex structure of hydrogen column density that is extended from the vicinity of the central engine to a more distant region. The hydrogen cloud, comparatively  near the central region, exhibits partial ionization and undergoes fast temporal changes. This cloud disappeared after 2015 and has not reappeared since. On the other hand, another portion of the hydrogen cloud, situated far from the central X-ray emitting region, remained relatively stable over time. It is important to note that the X-ray photons are not directly affected by the torus since the torus opening angle is significantly larger (approximately twice) than the inclination angle.

\subsection{Correlation between different parameters}
\begin{table}.
	\centering
	\caption{Correlation between parameters, obtained from the spectral fitting of data from observations with various X-ray observatories over a period of 22 years. Parameter-1 and Parameter-2, shown in the first and second columns, respectively, are involved in the correlation study. The Pearson Correlation Coefficient (PCC) is shown in the third column, and the corresponding $p$-values are shown in the fourth column. A negative PCC value shows an anti-correlation between the two parameters, whereas the opposite is valid for a positive PCC value.}
	\label{tab:par_cor}
	\begin{tabular}{l c c c } % four columns, alignment for each
     \hline
Parameter-1 & Parameter-2 & PCC & $p-value$ \\
\hline
$\Gamma$ &$\rm log\lambda_{\rm Edd}$ & +0.42&  0.09\\
$\Gamma$ &$\rm logL_{x}$ &+0.41& $ 0.10$\\
$\rm logL_{x}$ &$N_{\rm H1}$ &  -0.32 & 0.21\\
$N_{\rm H1}$ & $N_{\rm H2}$ & +0.64 &$ 0.24$ \\
$N_{\rm H2}$ &$C^{\xi}_{f}$&  +0.87 & 0.05\\
\hline
	\end{tabular}
\end{table}

\begin{figure} 
	\centering
	\includegraphics[scale=0.7]{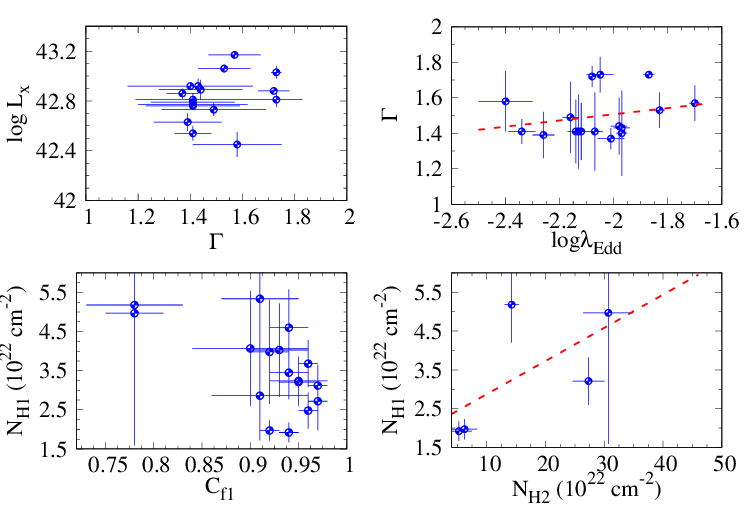}
	\caption{Correlation between different spectral parameters. }
	\label{fig:corr} 
\end{figure}

Our spectral analysis provides a meaningful understanding of the different properties of the X-ray emitting regions in Mrk~6. It is observed that the source exhibited various spectral states, with the photon index $\Gamma$ ranging from $1.37\pm0.12$ to $1.73\pm0.12$. We have also examined the correlation among various spectral parameters. Selected correlations between different parameters are presented in Figure~\ref{fig:corr}. We utilize the Pearson Correlation Coefficient (PCC\footnote{\url{https://www.socscistatistics.com/tests/pearson/default2.aspx}}) to check the order of correlations between different spectral parameters.  We have observed a weak correlation between the photon index $(\Gamma)$ and $\log \lambda_{\rm Edd}$ with the Pearson correlation coefficient (PCC) of +0.42 $(p-value=0.09)$ (presented in Figure~\ref{fig:corr}). It is worth mentioning that the parameter $\lambda_{\rm Edd}$ is intricately represented by the accretion rate \citep[e.g.,][]{Done2012} in astrophysical accreting systems. As the accretion rate increases, the supply of soft photons increases. Consequently, the power-law index steepens, leading to the correlation observed between $(\Gamma)$ and $\log \lambda_{\rm Edd}$. An increase in the supply of soft photons can produce more hard photons by interacting with the Compton cloud. However, this process also leads to cooling and shrinking of the Compton cloud. As a result, the X-ray luminosity of the continuum does not increase proportionately. Therefore, we do not find any significant correlation between $(\Gamma)$ and the 3.0-10.0 keV continuum luminosity $(L_x)$ (presented in Figure~\ref{fig:corr}) as the correlation coefficient PCC=+0.41 with $p-value=0.10$.

On examining the ambient medium surrounding the X-ray emitting region and the column densities along the line of sight, it is generally observed that the X-ray luminosity decreases with an increase in the hydrogen column density $(N_{\rm H1})$ along the line of sight. As the luminosity is an intrinsic source property and does not depend on absorption due to matter along the line of sight, we do not find any strong correlation or anti-correlation between $L_x$ and $N_{\rm H1}$.  The correlation coefficient is found to be PCC=-0.32 with a $(p-value=0.21)$. This suggests that the nature of the source remained unchanged, whereas there was a change in the density of the hydrogen cloud. As the column density increases, the amount of hydrogen gas in the ambient medium around the X-ray emitting region is expected to increase.  Consequently, we find a correlation between $N_{\rm H1}$ and $N_{\rm H2}$, with a PCC of +0.64 $(p-value=0.24)$. This indicates that as the gas density rises in the ambient medium around the central engine, it covers a larger area.  This behaviour is reflected in Mrk~6, where we find a strong correlation between $N_{\rm H2}$ and the ionized gas covering factor $C^{\xi}_{\rm f}$, with a PCC of +0.87 $(p-value=0.05)$.

In this study, we explore the correlation between different parameters and find that the variation of these parameters can be explained from the physical point of view.  However, it is important to note that the p-value associated with some correlation is greater than 0.05, indicating that the correlation may not be statistically significant at the $5\%$ level. This suggests that the observed relationship could potentially be due to random variability. Further investigation or a larger sample size may be needed to draw more accurate conclusions. Photon index $(\Gamma)$, which is directly measured from a simple power-law fitting to the X-ray spectrum, is related to other physical quantities such as X-ray continuum luminosity $(L_x)$ and/or Eddington ratio $(\lambda_{\rm Edd})$ of the source. The neutral hydrogen column density along the line of sight $(N_{\rm H1})$ is also found to be non-variable and correlated with the ionized hydrogen column density $(N_{\rm H2})$ of this source. These correlations and anti-correlations between different parameters provide valuable insights into the behavior of the X-ray emitting region in Mrk~6 and are consistent with known trends observed in other AGNs. We successfully explain their behavior using the correlations and anti-correlations between them. This result provides a better understanding of the complex behavior of Mrk~6.

\section{Conclusions}
\label{sec:conclusion}
We conducted a detailed temporal and spectral analysis of the X-ray observations of Mrk~6 over a time period of $\sim 22$ years $(2001-2022)$. During this observation period, we observe a minor variation in the various spectral parameters of this source, such as photon index, X-ray luminosity, and Eddington ratio. We explored the nature of obscuring materials of this source and tried to understand its complex nature. Furthermore, our analysis of the broadband spectrum unveiled insights into the properties of the reflecting medium, allowing us to estimate the intrinsic parameters like the inclination angle and torus opening angle. In this summary, we present an outline of our key findings.
\begin{enumerate}

    \item[1.] 
    
    Although Mrk~6 displays characteristics of a changing-look AGN (CLAGN) from optical observation in the past, our X-ray spectral analysis shows a  marginal variation in the X-ray luminosity and Eddington ratio over a period of 22 years. This indicates that in the X-ray regime, the source did not show any significant change in its behavior during the observational period. It is noted that the X-ray continuum luminosity varies between $3\times10^{42}$ \ergsec~to  $15\times10^{42}$\ergsec, and the corresponding Eddington ratio changes from 0.004 to 0.020 in the 22 years of the observational period. Based on the calculated Eddington ratio, it is evident that the source remains in a sub-Eddington regime during these observations.

     \item[2.]   We observed a complex variable structure of the obscuring absorber of Mrk~6. The hydrogen cloud, relatively closer to the central engine, appears partially ionized and extends up to NLR or the Torus region. This section of the cloud displays complex variability and undergoes rapid temporal changes. We observed this component of the hydrogen cloud up to the 2015 observations and predicted that it would have disappeared between 2015 and 2019. In contrast, another portion, situated far from the central engine, remained relatively stable over time. It is important to note that the torus did not have a direct impact on the X-ray observations, as the torus opening angle is significantly larger than the inclination angle.

  \item[3.] 
  Correlation study of light curves in soft (0.5-3.0 keV range) and hard (3.0-10.0 keV range) bands yielded a fairly strong correlation with approximately zero delay for long-term observations. The detection of this correlation suggests that the photons in both energy bands originate from the same physical mechanism. However, in the shorter time scale (a few ks), we observed only a weak correlation or no correlation between these two energy bands. On the other hand, in the high energy band (above 10 keV), we did not notice any significant correlations with different energy bands. It is important to note that a strong correlation between the soft and hard bands exists when the structure of column density is relatively simpler.
  
 \item[4.] 
  
   From the temporal study, we report that the fractional rms amplitude ($F_{\rm var}$) of the source is below $10\%$ for the shorter timescale $(\sim$60 ks) and above $20\%$ for longer timescale ($\sim$ weeks).

\end{enumerate}

\section*{Acknowledgements}

We sincerely thank the anonymous referee for his/her insightful comments and constructive suggestions that helped us to improve the manuscript. The research work at the Physical Research Laboratory, Ahmedabad, is funded by the Department of Space, Government of India. The data and/or software used for this work is taken from the High Energy Astrophysics Science Archive Research Center (HEASARC), which is a service of the Astrophysics Science Division at NASA/GSFC and the High Energy Astrophysics Division of the Smithsonian Astrophysical Observatory. This work has made use of data obtained from the \nustar~ mission, a project led by Caltech, funded by NASA, and managed by NASA/JPL, and has utilized the NuSTARDAS software package, jointly developed by the ASDC, Italy, and Caltech, USA. This work has used data from the \suz, a collaborative mission between the space agencies of Japan (JAXA) and the USA (NASA). This work made use of data \Swift~supplied by the UK Swift Science Data Centre at the University of Leicester. This research has made use of observations obtained with \xmm~, an ESA science mission with instruments and contributions directly funded by ESA Member States and NASA.

\section*{Data Avilability}
We used archival data of {\it Swift}/XRT, \xmm, {\it Suzaku} and {\it NuSTAR} observatories for this work. These data are publicly available on their corresponding websites. Appropriate links are given in the text.



\bibliographystyle{mnras}
\bibliography{ref} % if your bibtex file is called example.bib

%\begin{thebibliography}{99}
%\end{thebibliography}

%%%%%%%%%%%%%%%%%%%%%%%%%%%%%%%%%%%%%%%%%%%%%%%%%%

%%%%%%%%%%%%%%%%% APPENDICES %%%%%%%%%%%%%%%%%%%%%

\appendix

%%%%%%%%%%%%%%%%%%%%%%%%%%%%%%%%%%%%%%%%%%%%%%%%%%

% Don't change these lines
\bsp	% typesetting comment
\label{lastpage}
\end{document}